\documentclass[journal,twocolumn]{IEEEtran}
\usepackage{amsfonts}
\usepackage{amsmath}
\usepackage{amssymb}
\usepackage{algorithm}
\usepackage{algorithmic}
\usepackage{bm}
\usepackage{cite}
\usepackage{color}
\usepackage{dsfont}
\usepackage{enumerate}
\usepackage{epsfig}
\usepackage{epstopdf}
\usepackage{graphicx}
\usepackage{mathrsfs}
\usepackage{mathtools}
\usepackage{relsize}

\def\BD{\begin{displaymath}}
\def\BE{\begin{equation}}
\def\BEA{\begin{eqnarray}}
\def\BEAs{\begin{eqnarray*}}
\def\ED{\end{displaymath}}
\def\EE{\end{equation}}
\def\EEA{\end{eqnarray}}
\def\EEAs{\end{eqnarray*}}

\def\bA{{\bf A}}

\def\bB{{\bf B}}

\def\be{{\bf e}}
\def\bF{{\bf F}}
\def\bG{{\bf G}}

\def\bH{{\bf H}}

\def\bI{{\bf I}}

\def\bL{{\bf L}}

\def\bn{{\bf n}}
\def\bP{{\bf P}}

\def\bQ{{\bf Q}}

\def\bR{{\bf R}}
\def\br{{\bf r}}

\def\bs{{\bf s}}

\def\bU{{\bf U}}
\def\bu{{\bf u}}
\def\bV{{\bf V}}

\def\bW{{\bf W}}
\def\bw{{\bf w}}
\def\bX{{\bf X}}
\def\bx{{\bf x}}
\def\bY{{\bf Y}}
\def\by{{\bf y}}
\def\bZ{{\bf Z}}
\def\bz{{\bf z}}

\def\b_eta{\mbox{\boldmath $\eta$}}

\def\bnu{\mbox{\boldmath $\nu$}}

\def\diag { {\rm diag} }

\def\trace{ {\rm tr} }

\def\L2{L^{2}[-{\pi \over 2} , {\pi \over 2}]}

\def\bnu{\bnu}

\DeclarePairedDelimiterX\set[1]\lbrace\rbrace{\def\given{\;\delimsize\vert\;}#1}
\graphicspath{{./}{Figures/}}

\newtheorem{theorem}{Theorem}

\newtheorem{proposition}{Proposition}

\newenvironment{example}[1][Example]{\begin{trivlist}
\item[\hskip \labelsep {\bfseries #1}]}{\end{trivlist}}

\begin{document}
\title{Generalized Quadratic Matrix Programming:
A Unified Framework for Linear Precoding With Arbitrary Input Distributions}

\author{Juening~Jin, Yahong~Rosa~Zheng,~\IEEEmembership{Fellow,~IEEE,} ~Wen~Chen,~\IEEEmembership{Senior Member,~IEEE,} ~Chengshan~Xiao,~\IEEEmembership{Fellow,~IEEE}

\thanks{The work of Y. R. Zheng and C. Xiao was supported in part by US National Science Foundation under Grants ECCS-1231848, ECCS-1408316 and ECCS-1539316. The work of W. Chen was supported in part by the national 973 project under Grant 2012CB316106 and the national 863 project under Grant 2015AA01A710. This work has been carried out while J. Jin is visiting Missouri University of Science and Technology. Part of the material in this paper was presented at the IEEE GLOBECOM, Washington, DC, 2016.}
\thanks{J. Jin is with the Department of Electronic Engineering, Shanghai Jiao Tong University, Shanghai 200240, China (E-mail:
jueningjin@gmail.com).}
\thanks{Y. R. Zheng and C. Xiao are with the Department of Electrical and Computer Engineering,
 Missouri University of Science and Technology, Rolla, MO  65409, USA
 (E-mail: zhengyr@mst.edu; xiaoc@mst.edu).}
\thanks{W. Chen is with the Shanghai Key Laboratory of Navigation and Location
Based Services, Shanghai Jiao Tong University, Shanghai 200240, China,
and also with the School of Electronics Engineering and Automation, Guilin
University of Electronics Technology, Guilin 541004, China (E-mail: wenchen@sjtu.edu.cn).}
}


\maketitle
\begin{abstract}
This paper investigates a new class of non-convex optimization, which provides a unified framework for linear precoding in single/multi-user multiple-input multiple-output (MIMO) channels with arbitrary input distributions. The new optimization is called generalized quadratic matrix programming (GQMP). Due to the nondeterministic polynomial time (NP)-hardness of GQMP problems, instead of seeking globally optimal solutions, we propose an efficient algorithm which is guaranteed to converge to a Karush-Kuhn-Tucker (KKT) point. The idea behind this algorithm is to construct explicit concave lower bounds for non-convex objective and constraint functions, and then solve a sequence of concave maximization problems until convergence. In terms of application, we consider a downlink underlay secure cognitive radio (CR) network, where each node has multiple antennas. We design linear precoders to maximize the average secrecy (sum) rate with finite-alphabet inputs and statistical channel state information (CSI) at the transmitter. The precoding problems under secure multicast/broadcast scenarios are GQMP problems, and thus they can be solved efficiently by our proposed algorithm. Several numerical examples are provided to show the efficacy of our algorithm.
\end{abstract}
\begin{IEEEkeywords}
Generalized quadratic matrix programming, non-convex optimization, MIMO, linear precoding, secrecy sum rate maximization, arbitrary input distributions.
\end{IEEEkeywords}

\IEEEpeerreviewmaketitle

\section{Introduction}
Optimization has been widely used in communications and signal processing. In many situations, the design and analysis of communication networks, when converted into mathematical forms, become certain types of optimization problems. However, finding the optimal solution to a general optimization problem is far from trivial. It is widely believed that the ``watershed'' in optimization is between convex and non-convex problems. Specifically, for any convex problem, the ellipsoid algorithm \cite{ben2001lectures} can be used to get a global optimum with arbitrary precision, and its complexity is a polynomial function with the problem size. In contrast, non-convex problems are generally nondeterministic polynomial time (NP)-hard, which implies that there exists no polynomial time algorithm that can solve general non-convex problems to global optimality unless the complexity classes P and NP are proven to be equal.

Although it is challenging to handle non-convex optimization, much progress has been made for certain types of non-convex problems, such as difference-of-convex (DC) programming, quadratic constrained quadratic programming (QCQP) and signomial programming (SP), by means of convex optimization approaches\cite{tao2005dc,yuille2003concave,lipp2016variations,quoc2011sequential,luo2010semidefinite,mehanna2015feasible,chiang2007power,lange2014mm}. In \cite{tao2005dc}, the authors revisited a DC algorithm that can address DC problems, whose objective and constraint functions are the difference of two convex functions (not necessarily differentiable). When both objective and constraints of a DC problem are differentiable, the DC algorithm in \cite{tao2005dc} becomes another algorithm called convex-concave procedure \cite{yuille2003concave,lipp2016variations}. The work in \cite{luo2010semidefinite} introduced the semidefinite relaxation (SDR) technique for non-convex QCQP problems. The main idea of SDR is to lift QCQP problems into the positive semidefinite matrix space, and then relax the non-convex rank one constraint. This technique is very powerful especially when the problem size is small. Moreover, a feasible point pursuit successive convex approximation was proposed in \cite{mehanna2015feasible} for nonconvex QCQP problems. Finally, references \cite{chiang2007power,lange2014mm} proposed two different numerical algorithms, which are derived from the majorize-minimization (MM) framework \cite{hunter2004tutorial}, for non-convex SP problems.

However, many non-convex problems in communications and signal processing cannot be cast as DC, QCQP or SP problems. An important example is the linear precoder design for mutual information maximization in single-user multiple-input multiple-output (MIMO) Gaussian channels with finite-alphabet inputs \cite{palomar2006gradient,lozano2006optimum,xiao2008mutual,perez2010mimo,xiao2011globally}. It was revealed in \cite{xiao2011globally} that the input-output mutual information $\mathcal{I}(\bP^\mathrm{H}\bH^\mathrm{H}\bH\bP)$ is the composition of a concave function $\mathcal{I}(\bW)$ and a quadratic matrix function $\bW\!=\!\bP^\mathrm{H}\bH^\mathrm{H}\bH\bP$, where $\bP$ is the precoding matrix. Such a composite function is neither convex nor concave with respect to $\bP$, and it cannot be expressed as a DC, quadratic, or signomial function. For multiuser MIMO channels, including \cite{wu2012linearw,wu2012linearb,zeng2012globally,wu2013linear,juening2015,jin2016generalized}, precoding problems under finite-alphabet inputs are even more difficult, because they are generalizations of the single-user case. To the best of our knowledge, there does not exist any specific class of non-convex optimization that can capture the underlying structure of linear precoding for different communication channels and arbitrary channel input distributions.

\subsection{Contributions}
The contributions of this paper are listed as follows:

First, we study a new class of non-convex optimization, which provides a unified framework for linear precoding under different MIMO channels and arbitrary input distributions. The new optimization is a generalization of the quadratic matrix programming \cite{beck2007quadratic}, and we call it generalized quadratic matrix programming (GQMP). A GQMP problem is defined as maximizing a generalized quadratic matrix function subject to generalized quadratic matrix inequality constraints, with both objective and constraints being non-convex functions. In this paper, we develop a numerical algorithm to solve GQMP problems efficiently. The solution obtained by our proposed GQMP algorithm reaches the Karsuh-Kuhn-Tucker (KKT) point. The key idea of this algorithm is to construct a concave lower bound for any generalized quadratic matrix function, and then replace the non-convex objective and constraint functions with the corresponding concave lower bounds. Subsequently, we solve a sequence of concave maximization problems until convergence. We further analyze the computational complexity of the GQMP algorithm and discuss two non-smooth generalizations of standard GQMP problems.

Second, we consider a downlink underlay secure cognitive radio (CR) network where a secondary-user transmitter (ST) communicates with $I$ secondary-user receivers (SRs) in the presence of $J$ eavesdroppers (EDs) and subject to interference threshold constraints at $K$ primary-user receivers (PRs). Each node in the network is equipped with multiple antennas. We address the fundamental problem of maximizing the average secrecy (sum) rate of secondary users through linear precoding under the following assumptions: 1) The ST employs finite-alphabet modulation schemes; 2) The ST only has the knowledge of statistical channel state information (CSI) of each network node. The linear precoding problems under both secure multicast and secure broadcast scenarios are GQMP problems, thus they can be solved efficiently by our proposed GQMP algorithm. Finally, we present several numerical results to evaluate the performance of the proposed precoding with different system parameters. These results show that when considering finite-alphabet systems, our proposed precoding significantly outperforms the conventional Gaussian precoding design.

\subsection{Notations}
The following notations are adopted throughout the paper: Boldface lowercase letters, boldface uppercase letters, and calligraphic letters are used to denote vectors, matrices and sets, respectively. The real and complex number fields are denoted by $\mathds{R}$ and $\mathds{C}$, respectively. The space of Hermitian $n\times n$ matrices is denoted by $\mathds{H}^n$. The superscripts $(\cdot)^{\mathrm{T}}$, $(\cdot)^{*}$ and $(\cdot)^\mathrm{H}$ stand for transpose, conjugate, and conjugate transpose operations, respectively. $\trace(\cdot)$ is the trace of a matrix; $[\cdot]^+$ denotes $\max(\cdot,0)$; $\mathbf{dom}(\cdot)$ denotes the domain of a function; $\bA^{\!\scriptscriptstyle(+)}$ denotes the positive definite part of a Hermitian matrix $\bA$, i.e., $\bA^{\!\scriptscriptstyle(+)}=\sum_{\lambda_i>0}\lambda_i\bu_i\bu_i^\mathrm{H}$, where $\lambda_i$ is the $i$-th eigenvalue of $\bA$, and $\bu_i$ is the corresponding eigenvector of $\bA$; $\bA^{\!\scriptscriptstyle(-)}$ denotes the negative definite part of $\bA$, i.e., $\bA^{\!\scriptscriptstyle(-)}=\sum_{\lambda_i<0}\lambda_i\bu_i\bu_i^\mathrm{H}$; $\|\!\cdot\!\|$ denotes the Euclidean norm of a vector; $E_{\bx}(\cdot)$ represents the statistical expectation with respect to $\bx$; $\bI$ and $\bm{0}$ denote an identity matrix and a zero matrix, respectively, with appropriate dimensions; $\bA\!\succeq
\!\bB$ represents $\bA-\bB$ is positive semidefinite; $\mathcal{I}(\cdot)$ represents the mutual information; $\log(\cdot)$ and $\ln(\cdot)$ are used for the base two logarithm and natural logarithm, respectively.

The rest of this paper is organized as follows. section II introduces and solves generalized quadratic matrix programming problems. Section III  sets up the network model and formulates the precoding problems. Section IV solves the linear precoding problems by generalized quadratic matrix programming. Section IV presents several numerical results and Section V draws the conclusion.

\section{Generalized Quadratic Matrix Programming}
A real-valued function $h(\bX)$ is said to be a composite quadratic matrix function if $h(\bX)$ can be expressed in the form
\begin{align}
h(\bX)=g(\bX^\mathrm{H}\bA\bX), \quad g\in \mathcal{G}
\end{align}
where $\bX\!\in\!\mathds{C}^{n \times r}$, $\bA\!\in\!\mathds{H}^n$, $g(\bW)\!:\! \mathds{H}^r\rightarrow \mathds{R}$, and $\mathcal{G}$ is the family of differentiable convex functions satisfying either matrix nondecreasing (MND) or matrix nonincreasing (MNI) condition.  The definition of MND and MNI are given respectively as \cite[ch. 3.6.1]{boyd2004convex}:
\begin{align}
\mathrm{MND}&:\;\mathrm{if} \; \bW_1\succeq \bW_2,\;g(\bW_1)\geq g(\bW_2)\label{MI}\\
\mathrm{MNI}&:\;\mathrm{if} \; \bW_1\succeq \bW_2,\;g(\bW_1)\leq g(\bW_2).\label{MD}
\end{align}
A linear combination of composite quadratic matrix functions is called a generalized quadratic matrix function
\begin{align}\label{GQMF0}
f(\bX)=\sum_{k=1}^{K}\alpha_{k}g_k(\bX^\mathrm{H}\bA_k\bX)
\end{align}
where $\alpha_{k}\!\in\!\mathds{R}$, $k=1,2,...,K$, $\bA_k\!\in\!\mathds{H}^n$, $k=1,2,...,K$ and $g_k\!\in\!\mathcal{G}$, $k=1,2,...,K$.

Programming problems dealing with generalized quadratic matrix functions are called generalized quadratic matrix programming (GQMP) problems. The standard form of a GQMP problem considered in this paper is given by
\begin{equation}\label{GQMP}
\begin{aligned}
& \underset{\bX\in \mathcal{X}}{\mathrm{maximize}}
& & f_0(\bX)\\
& \mathrm{subject \;to}
& & f_j(\bX)\geq 0, \; j=1,2,...,J
\end{aligned}
\end{equation}
where $\bX\!\in\!\mathds{C}^{n \times r}$, $\mathcal{X}$ is a compact convex set, and $f_j(\bX)$, $j=0,1,...,J$ are generalized quadratic matrix functions
\begin{align}\label{GQMF}
f_j(\bX)=\sum_{k=1}^{K_j}\alpha_{jk}g_{jk}(\bX^\mathrm{H}\bA_{jk}\bX).
\end{align}
Here we implicitly assume that the domain of problem \eqref{GQMP} is an open set containing $\mathcal{X}$, i.e.,
\begin{align}
\mathcal{X}\subset\mathbf{dom}f_0 \cap \mathbf{dom}f_1 \cap ... \cap \mathbf{dom}f_{\scriptscriptstyle J}.
\end{align}
This assumption ensures 1) $f_j(\bX)$, $j\!=\!0,1,...,J$ are differentiable at every $\bX\!\in\! \mathcal{X}$; 2) the domain $\bigcap_j \mathbf{dom}f_j$ has no effect on the optimal solution of problem \eqref{GQMP}.

Some major properties of the GQMP problem \eqref{GQMP} are listed as follows:
\begin{enumerate}[1)]
\item The GQMP problem can be expressed in the minimization form, since this is equivalent to the maximization of $-f_0(\bX)$, which is again a generalized quadratic matrix function. Similarly, the constraints of \eqref{GQMP} can be expressed as $f_j(\bX)\leq 0, \; j=1,2,...,J$.
\item The GQMP problem is a purely non-convex optimization problem, because generalized quadratic matrix functions $f_j(\bX)$, $j=0,1,...,J$ are non-concave with respect to $\bX$. When $J=0$, the simplified problem $\underset{\bX\in \mathcal{X}}{\mathrm{maximize}} \; f_0(\bX)$ is still a non-convex optimization problem.
\item When $\mathcal{X}\!=\!\mathds{C}^{n \times r}$, and $g_{jk}(\bW)\!=\!\trace(\bW)\!-\!c_{jk}$ for all $(j,k)$, the GQMP problem is reduced to a non-convex quadratic matrix programming problem \cite{beck2007quadratic}
    \begin{equation}\label{QMP}
    \begin{aligned}
    & \underset{\bX\in\mathds{C}^{n \times r}}{\mathrm{maximize}}
    & & \trace(\bX^\mathrm{H}\tilde{\bA}_0\bX)-\tilde{c}_0\\
    & \mathrm{subject \;to}
    & & \trace(\bX^\mathrm{H}\tilde{\bA}_j\bX)-\tilde{c}_j\geq 0, \; j=1,2,...,J
    \end{aligned}
    \end{equation}
    where $\tilde{\bA}_j\!=\!\sum_{k=1}^{\scriptscriptstyle K_{\!j}}\alpha_{jk}\bA_{jk}$, $j=0,1,...,J$, and $\tilde{c}_j\!=\!\sum_{k=1}^{\scriptscriptstyle K_{\!j}} \alpha_{jk}c_{jk}$, $j=0,1,...,J$. Problem \eqref{QMP} can be solved by the SDR technique, i.e., define $\bQ\!=\!\bX\bX^\mathrm{H}$ and then relax the non-convex constraint $\mathrm{rank}(\bQ)\leq \min\{n,r\}$.
\item The GQMP problem belongs to the class of NP-hard problems, since the quadratic matrix programming problem in \eqref{QMP} is NP-hard in general.
\end{enumerate}

Before stating our main results, we introduce a few definitions on the complex derivative and gradient. For a univariate function $f(x)\!:\! \mathds{C}\rightarrow \mathds{R}$, the definition of the complex derivative is given in \cite{hjorungnes2007complex}:
\begin{align}
\frac{\partial f}{\partial x^*}\triangleq \frac{1}{2} \bigg(\frac{\partial f}{\partial \Re(x)}+j\frac{\partial f}{\partial \Im(x)}\bigg)
\end{align}
where $\Re(\cdot)$ and $\Im(\cdot)$ are the real and image parts of a complex variable, respectively. For a multivariate function $f(\bX)\!:\! \mathds{C}^{n \times r}\rightarrow \mathds{R}$, the complex gradient matrix $\nabla_{\!\scriptscriptstyle\bX}f(\bX)$ is defined as
\begin{align}
\nabla_{\!\scriptscriptstyle\bX}f(\bX)\triangleq \bigg[\frac{\partial f}{\partial \bX^*_{ij}}\bigg]
\end{align}
where $\bX_{ij}$ denotes the $(i,j)$-th element of $\bX$.

\subsection{Motivation}
GQMP has a wide variety of applications in communications. In this subsection, we discuss some typical single/multi-user MIMO Gaussian channels, for which linear precoding with arbitrary input distributions can be formulated as GQMP problems.
\subsubsection{Single-user MIMO Gaussian channels}
The single-user MIMO Gaussian channel is modeled as \cite{xiao2011globally}
\begin{align}
\by=\bH\bP\bx+\bn
\end{align}
where $\bH\!\in\!\mathds{C}^{n\times r}$ is the complex channel matrix, $\bP\!\in\!\mathds{C}^{r\times r}$ is the linear precoder, $\bn\!\in\!\mathds{C}^{n\times 1}$ is the independent and identically
distributed (i.i.d.) circularly symmetric complex Gaussian noise with zero-mean and unit-variance, and $\bx\!\in\!\mathds{C}^{r\times 1}$ is the arbitrarily distributed channel input signal with zero-mean and covariance $E_{\bx}\big[\bx\bx^\mathrm{H}\big]=\bI$.

In Theorem 1 of \cite{xiao2011globally}, the authors presented three properties of the input-output mutual information $\mathcal{I}(\bx;\by)$ with arbitrarily distributed $\bx$:
\begin{align}\label{PMI}
\begin{aligned}
& \mathcal{I}(\bx;\by) \;\mathrm{is \;a\; function \;of} \;\bW=\bP^\mathrm{H}\bH^\mathrm{H}\bH\bP\\
& \nabla_{\!\scriptscriptstyle\bW}\mathcal{I}(\bx;\by)=\mathbf{\Phi}\\
& \mathcal{I}(\bx;\by) \;\mathrm{is \;a \;concave \;function \;with \;respect \;to}\; \bW\\
\end{aligned}
\end{align}
where $\mathbf{\Phi}$ is known as the minimum mean square matrix, i.e.,
\begin{align}
\mathbf{\Phi}=E\big[(\bx-E[\bx|\by])(\bx-E[\bx|\by])^\mathrm{H}\big].
\end{align}
The first property shows that $\mathcal{I}(\bx;\by)$ is a function of $\bW$, thus it can be expressed as $\mathcal{I}(\bW)$. The second property guarantees that $\mathcal{I}(\bW)$ is MND because $\mathbf{\Phi}$ is a positive semidefinite matrix \cite[ch. 3.6.1]{boyd2004convex}. The third property implies that $-\mathcal{I}(\bW)$ is a differentiable convex function of $\bW$. Based on the definition in \eqref{GQMF0}, $\mathcal{I}(\bP^\mathrm{H}\bH^\mathrm{H}\bH\bP)$ is a generalized quadratic matrix function of $\bP$. Furthermore, since the feasible precoders with maximum transmit power $\gamma$ form a convex set $\big\{\bP|\trace(\bP^\mathrm{H}\bP)\leq \gamma\big\}$, the following mutual information maximization problem is a GQMP problem:
\begin{equation}\label{PPPA}
\begin{aligned}
& \underset{\bP}{\mathrm{maximize}}
& & \mathcal{I}(\bP^\mathrm{H}\bH^\mathrm{H}\bH\bP)\\
& \mathrm{subject \;to}
& & \trace(\bP^\mathrm{H}\bP)\leq \gamma.
\end{aligned}
\end{equation}

\begin{example}[Example 1]
If $\bx$ is complex Gaussian distributed, the input-output mutual information is given by
\begin{align}\label{GE}
\mathcal{I}(\bx;\by)=\log\det(\bI+\bP^\mathrm{H}\bH^\mathrm{H}\bH\bP).
\end{align}
Since $\mathcal{I}(\bW)\!=\!\log\det(\bI+\bW)$ is a concave and MND function with respect to $\bW$, \eqref{GE} is a generalized quadratic matrix function of $\bP$.
\end{example}

\begin{example}[Example 2]
If each element of $\bx$ is uniformly distributed from a $Q$-ary discrete constellation set, the input-output mutual information is given by \cite{xiao2011globally}
\begin{align}\label{FAE}
\mathcal{I}(\bx;\by)=r\log Q
\!-\!\frac{1}{Q^r}\sum_{m=1}^{Q^r} \!E_{\bn}\bigg\{\!\log\!\sum_{k=1}^{Q^r}e^{-d_{m,k}}\bigg\}.
\end{align}
where $d_{m,k}\!=\!\|\bH\bP(\bx_m-\bx_k)\!+\!\bn\|^2\!-\!\|\bn\|^2$. According to \eqref{PMI}, the mutual information expression in \eqref{FAE} is a generalized quadratic matrix function of $\bP$.
\end{example}

\subsubsection{MIMO Gaussian broadcast channels}
The MIMO Gaussian broadcast channel is modeled as \cite{wu2012linearb}
\begin{align}
\by_i=\bH_i\big(\mathsmaller{\sum}_{j=1}^{m}\bP_{\!j}\bx_j\big)+\bn_i, \quad i=1,2,...,m
\end{align}
where $\bH_i\!\in\!\mathds{C}^{n\times r}$ is the complex channel matrix for the $i$-th receiver, $\bn_i\!\in\!\mathds{C}^{n\times 1}$ is the i.i.d. circularly symmetric complex Gaussian noise with zero-mean and unit-variance; $\bP_{\!j}\!\in\!\mathds{C}^{r\times r}$ and $\bx_j\!\in\!\mathds{C}^{r\times 1}$ are the linear precoder and the channel input signal for the $j$-th receiver, respectively. We assume that $\{\bx_j\}_{1\leq j\leq m}$ are independent, and $\bx_j\!\in\!\mathds{C}^{r\times 1}$ is arbitrarily distributed with zero-mean and covariance $E_{\bx_j}\big[\bx_j\bx_j^\mathrm{H}\big]=\bI$.

The weighted sum-rate maximization and power minimization problems with linear precoding can be formulated respectively as
\begin{equation}\label{BC1}
\begin{aligned}
& \underset{\bP}{\mathrm{maximize}}
& & \sum_{i=1}^{m}\mu_iR_i(\bP)\\
& \mathrm{subject \;to}
& & \trace(\bP^\mathrm{H}\bP)\leq \gamma.
\end{aligned}
\end{equation}
\begin{equation}\label{BC2}
\begin{aligned}
\quad\;\;\;& \underset{\bP}{\mathrm{minimize}}
& & \trace(\bP^\mathrm{H}\bP)\\
& \mathrm{subject \;to}
& & R_i(\bP)\geq \bar{R}_i,\quad \forall i.
\end{aligned}
\end{equation}
where $\bP=\big[\bP_{\!1},\bP_{\!2},...,\bP_{\!m}\big]$, $\mu_i\geq 0$ with $\sum_i \mu_i=m$, $\!R_i(\bP)\!\!=\!\!\mathcal{I}(\bx_i;\by_i)\!$ represents the achievable rate for the $i$-th receiver, $\gamma$ is the maximum total transmit power, and $\bar{R}_i$ is the minimum rate requirement for the $i$-th receiver. Using the chain rule for mutual information \cite{cover2012elements}, $R_i(\bP)$ can be expressed alternatively as
\begin{align}
R_i(\bP)=\mathcal{I}\big(\{\bx_j\}_{1\leq j\leq m};\by_i\big)-\mathcal{I}\big(\{\bx_j\}_{j\neq i};\by_i|\bx_i\big).
\end{align}
According to \eqref{PMI}, $R_i(\bP)$ is a generalized quadratic matrix function because it is the difference of two composite quadratic matrix functions. Thus problems \eqref{BC1} and \eqref{BC2} belong to the class of GQMP problems.

\subsubsection{MIMO Gaussian interference channels}
The MIMO Gaussian interference channel is modeled as \cite{wu2013linear}
\begin{align}
\by_j\!=\!\bH_{jj}\bP_{\!j}\bx_j+\!\!\sum_{i=1,i\neq j}^{m}\!\!\bH_{ij}\bP_{\!i}\bx_i+\bn_j, \;\; j\!=\!1,2,...,m
\end{align}
where $\bH_{ij}\!\in\!\mathds{C}^{n\times r}$ is the complex channel matrix between the $i$-th transmitter and the $j$-th receiver, $\bn_i\!\in\!\mathds{C}^{n\times 1}$ is the i.i.d. circularly symmetric complex Gaussian noise with zero-mean and unit-variance; $\bP_{\!i}\!\in\!\mathds{C}^{r\times r}$ and $\bx_i\!\in\!\mathds{C}^{r\times 1}$ are the linear precoder and the channel input signal at the $i$-th transmitter, respectively. We assume that $\{\bx_i\}_{1\leq i\leq m}$ are independent, and $\bx_i\!\in\!\mathds{C}^{r\times 1}$ is arbitrarily distributed with zero-mean and covariance $E_{\bx_i}\big[\bx_i\bx_i^\mathrm{H}\big]=\bI$.

The weighted sum-rate maximization problem with linear precoding can be formulated as
\begin{equation}\label{IC}
\begin{aligned}
& \underset{\{\bP_{\!i}\}}{\mathrm{maximize}}
& & \sum_{j=1}^{m}\mu_jR_j(\{\bP_{\!i}\})\\
& \mathrm{subject \;to}
& & \trace(\bP_{\!i}^\mathrm{H}\bP_{\!i})\leq \gamma_i, \quad i=1,2,...,m.
\end{aligned}
\end{equation}
where $\mu_j\geq 0$ with $\sum_j \mu_j=m$, $\{\bP_{\!i}\}$ denotes the collection of all precoders $\{\bP_{\!1},\bP_{\!2},...,\bP_{\!m}\}$, $R_j(\{\bP_{\!i}\})=\mathcal{I}(\bx_j;\by_j)$ represents the achievable rate at the $j$-th transmitter, and $\gamma_i$ is the maximum transmit power for the $i$-th transmitter. Using the chain rule for mutual information, $R_j(\{\bP_{\!i}\})$ can be expressed alternatively as
\begin{align}
R_j(\{\bP_{\!i}\})=\mathcal{I}\big(\{\bx_i\}_{1\leq i\leq m};\by_j\big)-\mathcal{I}\big(\{\bx_i\}_{i\neq j};\by_j|\bx_j\big).
\end{align}
The achievable rate $R_j(\{\bP_{\!i}\})$ is the difference of two composite quadratic matrix functions with respect to $\bP\!=\!\diag\{\bP_{\!1},\bP_{\!2},...,\bP_{\!m}\}$, where $\diag\{\cdot\}$ represents a block diagonal matrix. Therefore, $R_j(\{\bP_{\!i}\})$ is a generalized quadratic matrix function of $\bP$, and problem \eqref{IC} belongs to the class of GQMP problems.

\subsubsection{MIMO Gaussian wiretap channels}
The MIMO Gaussian wiretap channel is modeled as \cite{wu2012linearw}
\begin{equation}
\begin{aligned}
&\by_\mathrm{r}=\bH_\mathrm{r}\bP\bx+\bn_\mathrm{r}\\
&\by_\mathrm{e}=\bH_\mathrm{e}\bP\bx+\bn_\mathrm{e}
\end{aligned}
\end{equation}
where $\by_\mathrm{r}\!\in\!\mathds{C}^{n\times 1}$ and $\by_\mathrm{e}\!\in\!\mathds{C}^{n\times 1}$ are received signals at the intended receiver and the eavesdropper, respectively; $\bH_\mathrm{r}\!\in\!\mathds{C}^{n\times r}$ and $\bH_\mathrm{e}\!\in\!\mathds{C}^{n\times r}$ are complex channel matrices; $\bP\!\in\!\mathds{C}^{r\times r}$ is the linear precoder at the transmitter; $\bn_\mathrm{r}\!\in\!\mathds{C}^{n\times 1}$ and $\bn_\mathrm{e}\!\in\!\mathds{C}^{n\times 1}$ are i.i.d. circularly symmetric complex Gaussian noises with zero-means and covariances $\sigma^2\bI$; and $\bx\!\in\!\mathds{C}^{r\times 1}$ is the arbitrarily distributed channel input signal with zero-mean and covariance $E_{\bx}\big[\bx\bx^\mathrm{H}\big]=\bI$.

The secrecy rate maximization problem with linear precoding can be formulated as
\begin{equation}\label{WC}
\begin{aligned}
& \underset{\bP}{\mathrm{maximize}}
& & \mathcal{I}(\bx;\by_\mathrm{r})-\mathcal{I}(\bx;\by_\mathrm{e})\\
& \mathrm{subject \;to}
& & \trace(\bP^\mathrm{H}\bP)\leq \gamma.
\end{aligned}
\end{equation}
where $\gamma$ is the maximum transmit power. Since $\mathcal{I}(\bx;\by_\mathrm{r})-\mathcal{I}(\bx;\by_\mathrm{e})$ is a generalized quadratic matrix function of $\bP$, problem \eqref{WC} belongs to the class of GQMP problems.

\subsection{Algorithm Design}
In this subsection, we design a numerical algorithm for GQMP problems by investigating the underlying structure of composite quadratic matrix functions $g(\bX^\mathrm{H}\bA\bX)$. For every $g(\bX^\mathrm{H}\bA\bX)$, we provide the corresponding concave lower bound $l(\bX;\bX_0)$ and convex upper bound $u(\bX;\bX_0)$, which depend on an arbitrary matrix $\bX_0\!\in\!\mathds{C}^{n \times r}$ and satisfy the following three conditions:
\begin{align}\label{COND}
& l(\bX;\bX_0)\leq g(\bX^\mathrm{H}\bA\bX)\leq u(\bX;\bX_0) \;\mathrm{for \;all}\; \bX\nonumber\\
& g(\bX^\mathrm{H}\bA\bX)=l(\bX;\bX_0)=u(\bX;\bX_0) \;\mathrm{when}\; \bX=\bX_0\\
& \nabla_{\!\scriptscriptstyle\bX} g(\bX^\mathrm{H}\bA\bX)=\nabla_{\!\scriptscriptstyle\bX} l(\bX;\bX_0)=\nabla_{\!\scriptscriptstyle\bX} u(\bX;\bX_0) \; \mathrm{when}\; \bX=\bX_0.\nonumber
\end{align}
In other words, $g(\bX^\mathrm{H}\bA\bX)$ lies between $l(\bX;\bX_0)$ and $u(\bX;\bX_0)$, and it is tangent to both $l(\bX;\bX_0)$ and $u(\bX;\bX_0)$ when $\bX\!=\!\bX_0$. The conditions in \eqref{COND} are necessary for us to design an ascent algorithm that converges to a KKT point of problem \eqref{GQMP}.

\begin{theorem}
The concave lower bound of $g(\bX^\mathrm{H}\bA\bX)$ is
\begin{align}
l(\bX;\bX_0)\!=\!\trace\big(\bL(\bX)^\mathrm{H}\bG\big)\!+\!g(\bX_0^\mathrm{H}\bA\bX_0)\!-\!\trace\big(\bX_0^\mathrm{H}\bA\bX_0\bG\big) \label{LB}
\end{align}
where $\bG\!\in\!\mathds{H}^r$ is the complex gradient of $g(\bW)$ at $\bW=\bX_0^\mathrm{H}\bA\bX_0$, i.e., $\bG\!=\!\nabla_{\!\scriptscriptstyle\bW}g(\bX_0^\mathrm{H}\bA\bX_0)$, and $\bL(\bX)$ is given by
\begin{align}
\bL(\bX)=\left\{
\begin{aligned}
&\bL_{1}, \quad g(\bW) \; \mathrm{is}\; \mathrm{MND}\\
&\bL_{2}, \quad g(\bW) \; \mathrm{is}\; \mathrm{MNI}
\end{aligned}
\right.
\end{align}
with $\bL_1\!=\!\bX^\mathrm{H}\bA^{\!\scriptscriptstyle(-)}\bX\!+\!\bX^\mathrm{H}\bA^{\!\scriptscriptstyle(+)}\bX_0\!+\!\bX_0^\mathrm{H}\bA^{\!\scriptscriptstyle(+)}\bX\!-\!\bX_0^\mathrm{H}\bA^{\!\scriptscriptstyle(+)}\bX_0$, $\bL_2\!=\!\bX^\mathrm{H}\bA^{\!\scriptscriptstyle(+)}\bX\!+\!\bX^\mathrm{H}\bA^{\!\scriptscriptstyle(-)}\bX_0\!+\!\bX_0^\mathrm{H}\bA^{\!\scriptscriptstyle(-)}\bX\!-\!\bX_0^\mathrm{H}\bA^{\!\scriptscriptstyle(-)}\bX_0$.
\end{theorem}
\begin{IEEEproof}
See Appendix A.
\end{IEEEproof}

\begin{theorem}
The convex upper bound of $g(\bX^\mathrm{H}\bA\bX)$ is
\begin{align}\label{UB}
u(\bX;\bX_0)\!=\!\left\{
\begin{aligned}
&g\big(\bU_1(\bX)\big), \quad g(\bW) \; \mathrm{is}\; \mathrm{MND}\\
&g\big(\bU_2(\bX)\big), \quad g(\bW) \; \mathrm{is}\; \mathrm{MNI}
\end{aligned}
\right.
\end{align}
with
\begin{align*}
&\bU_1(\bX)\!=\!\bX^\mathrm{H}\bA^{\!\scriptscriptstyle(+)}\bX\!+\!\bX^\mathrm{H}\bA^{\!\scriptscriptstyle(-)}\bX_0\!+\!\bX_0^\mathrm{H}\bA^{\!\scriptscriptstyle(-)}\bX\!-\!\bX_0^\mathrm{H}\bA^{\!\scriptscriptstyle(-)}\bX_0\\ &\bU_2(\bX)\!=\!\bX^\mathrm{H}\bA^{\!\scriptscriptstyle(-)}\bX\!+\!\bX^\mathrm{H}\bA^{\!\scriptscriptstyle(+)}\bX_0\!+\!\bX_0^\mathrm{H}\bA^{\!\scriptscriptstyle(+)}\bX\!-\!\bX_0^\mathrm{H}\bA^{\!\scriptscriptstyle(+)}\bX_0.
\end{align*}
\end{theorem}
\begin{IEEEproof}
See Appendix A.
\end{IEEEproof}

Based on Theorems 1 and 2, a concave lower bound of $f_j(\bX)$ in \eqref{GQMF} is given by
\begin{align}\label{CLB}
\bar{f}_j(\bX;\bX_0)\!=\!\!\!\!\sum_{\alpha_{\!jk}>0}\!\alpha_{jk} l_{jk}(\bX;\bX_0)\!+\!\!\!\sum_{\alpha_{\!jk}<0}\!\alpha_{jk} u_{jk}(\bX;\bX_0)
\end{align}
where $l_{jk}(\bX;\bX_0)$ and $u_{jk}(\bX;\bX_0)$ represent the lower and upper bounds of $g_{jk}(\bX^\mathrm{H}\bA_{jk}\bX)$, respectively. By replacing each of $f_j(\bX)$, $j=0,1,...,J$ with the corresponding concave lower bound $\bar{f}_j(\bX;\bX_0)$, we obtain the following concave maximization problem
\begin{equation}\label{GQMPLB}
\begin{aligned}
& \underset{\bX\in \mathcal{X}}{\mathrm{maximize}}
& & \bar{f}_0(\bX;\bX_0)\\
& \mathrm{subject \;to}
& & \bar{f}_j(\bX;\bX_0)\geq 0, \; j=1,2,...,J.
\end{aligned}
\end{equation}
Then the MM framework \cite{hunter2004tutorial} can be exploited to find a KKT point of problem \eqref{GQMP} through solving a sequence of problems \eqref{GQMPLB} with different $\bX_0$. In the first iteration, we solve \eqref{GQMPLB} at initial $\bX_0$, and the optimal solution is denoted by $\bX_{1}^{\mathrm{opt}}$. Then we replace $\bar{f}_j(\bX;\bX_0)$ in \eqref{GQMPLB} with $\bar{f}_j(\bX;\bX_1^{\mathrm{opt}})$, $j=0,1,...,J$, and solve problem \eqref{GQMPLB} again. At the $n$-th iteration, we solve \eqref{GQMPLB} by replacing $\bar{f}_j(\bX;\bX_0)$ with $\bar{f}_j(\bX;\bX_{n-1}^{\mathrm{opt}})$, $j=0,1,...,J$, where $\bX_{n-1}^{\mathrm{opt}}$ is the optimal solution of \eqref{GQMPLB} at the $(n-1)$-th iteration. The GQMP algorithm for solving problem \eqref{GQMP} is summarized in Algorithm 1.

\begin{algorithm}
\caption{: The GQMP algorithm}
\begin{algorithmic}
\STATE 1) Initialization: Given tolerance $\epsilon>0$, choose an initial \STATE \quad feasible point $\bX_0$, set $n\!=\!1$,
$s_{0}\!=\!f_0(\bX_0)$. Let $\bX_{n}^{\mathrm{opt}}$ re-\STATE \quad presents the optimal solution of \eqref{GQMPLB} at the $n$-th iteration.
\STATE 2) Stopping criterion: if $|s_{n}-s_{n-1}|>\epsilon$ go to the next step, \STATE \quad otherwise STOP.
\STATE 3) Concave approximation:
\STATE \quad a) replace $\bX_0$ in \eqref{GQMPLB} with $\bX_{n}^{\mathrm{opt}}$ and solve problem \eqref{GQMPLB} \STATE \quad\quad to obtain $\bX_{n+1}^{\mathrm{opt}}$.
\STATE \quad b) set $s_{n+1}=f_0(\bX_{n+1}^{\mathrm{opt}})$.
\STATE 4) Set $n:=n+1$ and go to step 2).
\STATE 5) Output: $\bX_{n}^{\mathrm{opt}}$.
\end{algorithmic}
\end{algorithm}
The convergence of Algorithm 1 is presented by the following proposition.
\begin{proposition}
Every limit point of the iterates $\{\bX_{n}^{\mathrm{opt}}\}$ generated by Algorithm 1 satisfies the KKT conditions of problem \eqref{GQMP}.
\end{proposition}
\begin{IEEEproof}
See Appendix B.
\end{IEEEproof}

Algorithm 1 is an iterative procedure between updating concave lower bounds and solving concave maximization problems \eqref{GQMPLB}. Since \eqref{GQMPLB} is a convex optimization problem, it can be solved efficiently by the interior-point method with Newton iterations. The total number of optimization variables in problem \eqref{GQMPLB} is $nr$, then the complexity order for solving \eqref{GQMPLB} with Newton iterations is about $\mathcal{O}((nr)^3)$ \cite{nocedal1999numerical}. Assuming that Algorithm 1 updates the concave lower bounds $T$ times, the overall complexity is then given by $\mathcal{O}(T(nr)^3)$.

\subsection{Non-smooth Generalization}
The GQMP algorithm developed in the last subsection can be used to handle non-smooth optimization problems. Herein, we discuss two possible non-smooth generalizations.
\subsubsection{GQMP with Min-rate Utility}
For a multiuser system, we often adopt a utility function to measure the overall system performance. Thus it is of interest to consider the following GQMP problem with the min-rate utility:
\begin{equation}\label{GQMPE}
\begin{aligned}
R=& \underset{\bX\in \mathcal{X}}{\mathrm{maximize}}
& & \min_{1\leq j\leq L}f_j(\bX)\\
& \mathrm{subject \;to}
& & f_j(\bX)\geq 0, \; j=L\!+\!1,2,...,J
\end{aligned}
\end{equation}
where $R$ is the optimal value of problem \eqref{GQMPE}, $\bX\!\in\!\mathds{C}^{n \times r}$, $\mathcal{X}$ is a compact convex set, and $f_j(\bX)$, $j=1,2,...,J$ are generalized quadratic matrix functions
\begin{align}
f_j(\bX)=\sum_{k=1}^{K_j}\alpha_{jk}g_{jk}(\bX^\mathrm{H}\bA_{jk}\bX).
\end{align}
Problem \eqref{GQMPE} is a non-smooth optimization problem, and our first step is to replace the non-smooth point-wise minimum operator with the log-sum-exp approximation via the following inequality \cite{chen2013markov,li2013transmit}
\begin{align}\label{LSI}
0\leq \min_{1\leq j\leq L} a_j-\frac{1}{\beta}\ln\sum_{j=1}^{L}e^{\beta a_j}\leq \frac{1}{|\beta|} \ln L, \quad \beta<0
\end{align}
which results in a smooth optimization problem
\begin{equation}\label{GQMPLB3}
\begin{aligned}
R_s(\beta)=\;& \underset{\bX\in \mathcal{X}}{\mathrm{maximize}}
& & \frac{1}{\beta}\ln\sum_{j=1}^{L}e^{\beta f_j(\bX)}\\
& \mathrm{subject \;to}
& &  f_j(\bX)\geq 0, \; j=L+1,2,...,J.
\end{aligned}
\end{equation}
Here $R_s(\beta)$ is the optimal value of \eqref{GQMPLB3}. The relationship between problems \eqref{GQMPE} and \eqref{GQMPLB3} is revealed as follows
\begin{align}
|R_s(\beta)-R|<\frac{1}{|\beta|} \ln L.
\end{align}
Then we can obtain a KKT point of \eqref{GQMPLB3} through solving a sequence of maximization problems
\begin{equation}\label{GQMPLB4}
\begin{aligned}
\;& \underset{\bX\in \mathcal{X}}{\mathrm{maximize}}
& & \frac{1}{\beta}\ln\sum_{j=1}^{L}e^{\beta\bar{f}_j(\bX;\bX_0)}\\
& \mathrm{subject \;to}
& &  \bar{f}_j(\bX;\bX_0)\geq 0, \; j=L+1,2,...,J
\end{aligned}
\end{equation}
where $\bar{f}_j(\bX;\bX_0)$ is the concave lower bound of $f_j(\bX)$, $j=1,2,...,J$. The objective function of problem \eqref{GQMPLB4} is concave because $-\ln\sum_j\exp(-f_j)$ is concave whenever $f_j$ is concave for all $j$. Therefore, problem \eqref{GQMPLB4} is a smooth concave maximization problem, whose optimal value can be readily attained. Subsequently, we invoke Algorithm 1 to solve \eqref{GQMPLB4} multiple times with sufficiently large $|\beta|$ until convergence, and a suboptimal solution of problem \eqref{GQMPE} is obtained.

\subsubsection{Secrecy Sum Rate Maximization}
Consider the secrecy sum rate maximization problem in multiuser multi-eavesdropper networks, where each node is equipped with multiple antennas. The secrecy sum rate in such networks has a very complicated form:
\begin{align}
\sum_{i=1}^{I}\!\min_{1\leq j\leq J} \big[R_{ij}\big]^{+}
\end{align}
where $\big[R_{ij}\big]^{+}$ represents the individual secrecy rate between the $i$-th user and the $j$-th eavesdropper.
Replacing each of $R_{ij}$ with a generalized quadratic matrix function, we obtain the following generalized GQMP problems:
\begin{equation}\label{GGQMP}
\begin{aligned}
& \underset{\bX\in \mathcal{X}}{\mathrm{maximize}}
& & \sum_{i=1}^{I}\!\min_{1\leq j\leq J} \big[f_{ij}(\bX)\big]^{+}
\end{aligned}
\end{equation}
where $\bX\!\in\!\mathds{C}^{n \times r}$, $\mathcal{X}$ is a compact convex set, and $f_{ij}(\bX)$, $\forall (i,j)$ are generalized quadratic matrix functions
\begin{align}
f_{ij}(\bX)=\sum_{k=1}^{K_{ij}}\alpha_{ijk}g_{ijk}(\bX^\mathrm{H}\bA_{ijk}\bX).
\end{align}

Problem \eqref{GGQMP} is extremely difficult to solve because of the non-concavity of $[\cdot]^+$. To see this, we first reformulate problem \eqref{GGQMP} as
\begin{equation}\label{GGQMP1}
\begin{aligned}
& \underset{\bX\in \mathcal{X}}{\mathrm{maximize}}
& & \sum_{i=1}^{I}\Big[\!\min_{1\leq j\leq J} f_{ij}(\bX)\Big]^{+}.
\end{aligned}
\end{equation}
Invoking Theorems 1 and 2, we can construct the concave lower bound of $f_{ij}(\bX)$, which is denoted by $\bar{f}_{ij}(\bX;\bX_0)$. Moreover, the point-wise minimum of concave functions is concave \cite{boyd2004convex}, i.e., $\min_j \bar{f}_{ij}(\bX;\bX_0)$ is a concave function. However, $\sum_{i}[\min_j \bar{f}_{ij}(\bX;\bX_0)]^{+}$ is neither concave nor quasi-concave, due to the non-concavity of $[\cdot]^+$. Therefore, Algorithm 1 cannot be applied directly to solve problem \eqref{GGQMP}.  The following proposition shows that a ``naive'' method can be used to solve problem \eqref{GGQMP} via solving $2^I\!-\!1$ GQMP problems:
\begin{equation}\label{S1231}
\begin{aligned}
&R_m\!=\underset{\bX\in \mathcal{X}}{\mathrm{maximize}}\;\;\sum_{i\in \mathcal{S}_m}\!\min_{1\leq j\leq J} f_{ij}(\bX)
\end{aligned}
\end{equation}
where $R_m$ is the optimal value of \eqref{S1231}, and $\mathcal{S}_m$ represents the $m$-th non-empty subset of $\{1,2,...,I\}$.
\begin{proposition}
Let $R$ denote the optimal value of problem \eqref{GGQMP}, then we have
\begin{align}
R=\max_{1\leq m\leq 2^{I}\!-\!1} R_m.
\end{align}
\end{proposition}
\begin{IEEEproof}
See Appendix B.
\end{IEEEproof}

Although Proposition 2 provides a trackable way to solve problem \eqref{GGQMP}, the complexity of this approach grows exponentially with respect to $I$. In the sequel, we design a more efficient algorithm for problem \eqref{GGQMP}.

Our first step is to reformulate problem \eqref{GGQMP1} as
\begin{equation}\label{GGQMP2}
\begin{aligned}
& \underset{\bX\in \mathcal{X}}{\mathrm{maximize}}
& & \sum_{i=1}^{I}\max_{\lambda_{i}\in [0,1]} \Big[\lambda_{i}\!\cdot\!\min_{1\leq j\leq J} f_{ij}(\bX)\Big].
\end{aligned}
\end{equation}
Problems \eqref{GGQMP1} and \eqref{GGQMP2} are equivalent because
\begin{align}
\Big[\!\min_{1\leq j\leq J} f_{ij}(\bX)\Big]^{+}=\max_{\lambda_{i}\in [0,1]} \Big[\lambda_{i}\!\cdot\!\min_{1\leq j\leq J} f_{ij}(\bX)\Big].
\end{align}
The optimal $\lambda_{i}$ of problem \eqref{GGQMP2} depends on $\bX$, and it has the following semi-closed form expression:
\begin{align}
\lambda_{i}^*(\bX)=\left\{
\begin{aligned}
&1, \quad \min_{1\leq j\leq J} f_{ij}(\bX)\geq 0\\
&0, \quad \min_{1\leq j\leq J} f_{ij}(\bX)< 0
\end{aligned}
\right., \quad i=1,2,...,I.
\end{align}
By plugging $\lambda_{i}^*(\bX)$ into \eqref{GGQMP2}, problem \eqref{GGQMP} is equivalent to the following optimization problem
\begin{equation}\label{GGQMP3}
\begin{aligned}
& \underset{\bX\in \mathcal{X}}{\mathrm{maximize}}
& & F(\bX)=\sum_{i=1}^{I}\Big[\lambda_{i}^*(\bX)\!\cdot\!\min_{1\leq j\leq J} f_{ij}(\bX)\Big].
\end{aligned}
\end{equation}
Note that for any $\bX_0\!\in\!\mathcal{X}$, the following inequalities hold
\begin{align}
F(\bX)&=\sum_{i=1}^{I}\max_{\lambda_{i}\in [0,1]} \Big[\lambda_{i}\!\cdot\!\min_{1\leq j\leq J} f_{ij}(\bX)\Big]\\
&\geq\sum_{i=1}^{I}\lambda_{i}^*(\bX_0)\!\cdot\!\min_{1\leq j\leq J} f_{ij}(\bX)\\
&\geq\sum_{i=1}^{I}\lambda_{i}^*(\bX_0)\!\cdot\!\min_{1\leq j\leq J} \bar{f}_{ij}(\bX;\bX_0). \label{LBF}
\end{align}
where $\bar{f}_{ij}(\bX;\bX_0)$ is the concave lower bound of $f_{ij}(\bX)$. Since $\lambda_{i}^*(\bX_0)\geq 0$, equation \eqref{LBF} serves as the concave lower bound of $F(\bX)$. Therefore, by replacing $F(\bX)$ with its concave lower bound in \eqref{LBF}, we obtain a concave maximization problem
\begin{equation}\label{GGQMP4}
\begin{aligned}
\underset{\bX\in \mathcal{X}}{\mathrm{maximize}}\;\;\bar{F}(\bX;\bX_0)\!=\!\!\sum_{i=1}^{I}\lambda_{i}^*(\bX_0)\!\cdot\!\!\min_{1\leq j\leq J} \!\bar{f}_{ij}(\bX;\bX_0).
\end{aligned}
\end{equation}
Using the log-sum-exp approximation in \eqref{LSI}, we can solve problem \eqref{GGQMP4} efficiently through solving a smooth convex problem. Then we can obtain a suboptimal solution of problem \eqref{GGQMP} by solving a sequence of problems \eqref{GGQMP4} with different $\bX_0$. The details are summarized in Algorithm 2, on the top of next page.
\begin{algorithm}
\caption{: The Generalized GQMP algorithm}
\begin{algorithmic}
\STATE 1) Initialization: Given tolerance $\epsilon>0$, choose an initial \STATE \quad feasible point $\bX_0$, set $n\!=\!1$,
$s_{0}\!=\!F(\bX_0)$. Let $\bX_{n}^{\mathrm{opt}}$ re-\STATE \quad presents the optimal solution of \eqref{GGQMP4} at the $n$-th iteration.
\STATE 2) Stopping criterion: if $|s_{n}-s_{n-1}|>\epsilon$ go to the next step, \STATE \quad otherwise STOP.
\STATE 3) Concave approximation:
\STATE \quad a) replace $\bX_0$ in \eqref{GGQMP4} with $\bX_{n}^{\mathrm{opt}}$ and solve problem \eqref{GGQMP4} \STATE \quad\quad to obtain $\bX_{n+1}^{\mathrm{opt}}$.
\STATE \quad b) set $s_{n+1}=F(\bX_{n+1}^{\mathrm{opt}})$.
\STATE 4) Set $n:=n+1$ and go to step 2).
\STATE 5) Output: $\bX_{n}^{\mathrm{opt}}$.
\end{algorithmic}
\end{algorithm}
The convergence of Algorithm 2 for problem \eqref{GGQMP} is guaranteed by the following proposition.

\begin{proposition}
Let $\bX_{n}^{\mathrm{opt}}$ represent the optimal solution of problem \eqref{GGQMP4} at the $n$-th iteration, then the sequence $\{F(\bX_{n}^{\mathrm{opt}})\}$ generated by Algorithm 2 converges.
\end{proposition}
\begin{IEEEproof}
See Appendix B.
\end{IEEEproof}

\section{System Model and Problem Formulation}
We consider a downlink underlay secure CR network depicted in Fig. 1. A secondary-user transmitter (ST) is serving $I$ secondary-user receivers (SRs) in the presence of $J$ eavesdroppers (EDs) and $K$ primary-user receivers (PRs). The channel output at the $i$-th SR, the $j$-th ED and the $k$-th PR are, respectively, given by
\begin{align}
\by_{i}&=\bH_{i}\bs+\bn_{h_{i}}, \quad i=1,2,...,I \nonumber \\
\bz_{j}&=\bG_{\!j}\bs+\bn_{g_{j}}, \quad j=1,2,...,J \nonumber \\
\bw_{k}&=\bF_{\!k}\bs+\bn_{f_{k}}, \quad k=1,2,...,K
\end{align}
where $\bH_{i}\!\in\!\mathds{C}^{N_{\!\scriptscriptstyle R}\times N_{\scriptscriptstyle T}}$, $\bG_{\!j}\!\in\!\mathds{C}^{N_{\!\scriptscriptstyle E}\times N_{\scriptscriptstyle T}}$ and $\bF_{\!k}\!\in\!\mathds{C}^{N_{\!\scriptscriptstyle P}\times N_{\scriptscriptstyle T}}$ are complex channel matrices from the ST to the $i$-th SR, the $j$-th ED, and the $k$-th PR, respectively; $\bs\!\in\!\mathds{C}^{N_{\scriptscriptstyle T}\times 1}$ is the channel input at the ST; $\bn_{h_{i}}\!\in\!\mathds{C}^{N_{\!\scriptscriptstyle R}\times 1}$, $\bn_{g_{j}}\!\in\!\mathds{C}^{N_{\!\scriptscriptstyle E}\times 1}$ and $\bn_{f_{k}}\!\in\!\mathds{C}^{N_{\!\scriptscriptstyle P}\times 1}$ are i.i.d. circularly symmetric complex Gaussian noises with zero-mean and unit-variance.

In this paper, we assume Kronecker correlation model, where the channel matrices can be written as \cite{xiao2004discrete}
\begin{align}\label{CMOD}
\bH_i&=\mathbf{\Phi}^{\frac{1}{2}}_{\!h_i}\tilde{\bH}_i\mathbf{\Theta}^{\frac{1}{2}}_{\!h_i}, \quad i=1,2,...,I \nonumber \\
\bG_{\!j}&=\mathbf{\Phi}^{\frac{1}{2}}_{\!g_j}\tilde{\bG}_j\mathbf{\Theta}^{\frac{1}{2}}_{\!g_j}, \quad j=1,2,...,J \nonumber \\
\bF_{\!k}&=\mathbf{\Phi}^{\frac{1}{2}}_{\!f_k}\tilde{\bF}_{\!k}\mathbf{\Theta}^{\frac{1}{2}}_{\!f_k}, \quad k=1,2,...,K.
\end{align}
Here $\tilde{\bH}_i\!\in\!\mathds{C}^{N_{\!\scriptscriptstyle R}\times N_{\scriptscriptstyle T}}$, $\tilde{\bG}_j\!\in\!\mathds{C}^{N_{\!\scriptscriptstyle E}\times N_{\scriptscriptstyle T}}$ and $\tilde{\bF}_{\!k}\!\in\!\mathds{C}^{N_{\!\scriptscriptstyle P}\times N_{\scriptscriptstyle T}}$ are random matrices with i.i.d. zero-mean unit-variance complex Gaussian entries; $\mathbf{\Theta}_{\!h_i}\!\in\!\mathds{C}^{N_{\scriptscriptstyle T}\times N_{\scriptscriptstyle T}}$, $\mathbf{\Theta}_{\!g_j}\!\in\!\mathds{C}^{N_{\scriptscriptstyle T}\times N_{\scriptscriptstyle T}}$ and $\mathbf{\Theta}_{\!f_k}\!\in\!\mathds{C}^{N_{\scriptscriptstyle T}\times N_{\scriptscriptstyle T}}$ are positive semidefinite transmit correlation matrices of $\bH_i$, $\bG_j$ and $\bF_{\!k}$, respectively. $\mathbf{\Phi}_{\!h_i}\!\in\!\mathds{C}^{N_{\scriptscriptstyle R}\times N_{\scriptscriptstyle R}}$, $\mathbf{\Phi}_{\!g_j}\!\in\!\mathds{C}^{N_{\scriptscriptstyle E}\times N_{\scriptscriptstyle E}}$ and $\mathbf{\Phi}_{\!f_k}\!\in\!\mathds{C}^{N_{\scriptscriptstyle P}\times N_{\scriptscriptstyle P}}$ are positive semidefinite receive correlation matrices of $\bH_i$, $\bG_j$ and $\bF_{\!k}$, respectively. We further assume that the receive antennas at SRs, EDs and PRs are uncorrelated, i.e., $\mathbf{\Phi}_{\!h_i}$, $\mathbf{\Phi}_{\!g_j}$ and $\mathbf{\Phi}_{\!f_k}$ are all identity matrices. Then the correlation model in \eqref{CMOD} becomes
\begin{align}
\bH_i&=\tilde{\bH}_i\mathbf{\Theta}^{\frac{1}{2}}_{\!h_i}, \quad i=1,2,...,I \nonumber \\
\bG_{\!j}&=\tilde{\bG}_j\mathbf{\Theta}^{\frac{1}{2}}_{\!g_j}, \quad j=1,2,...,J \nonumber \\
\bF_{\!k}&=\tilde{\bF}_{\!k}\mathbf{\Theta}^{\frac{1}{2}}_{\!f_k}, \quad k=1,2,...,K.
\end{align}
Note that the above uncorrelated assumption are only used to avoid cumbersome expressions in precoding design. As we will see in Section IV, even when $\mathbf{\Phi}_{\!h_i}$, $\mathbf{\Phi}_{\!g_j}$ and $\mathbf{\Phi}_{\!f_k}$ are
\begin{figure}[h]
  \setlength{\belowcaptionskip}{-0.4cm} 
  \centering
  \includegraphics[scale=0.38]{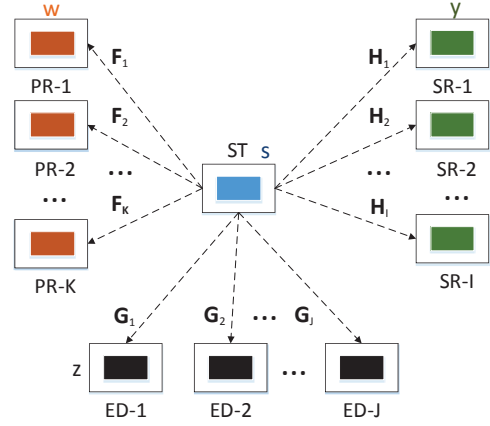}\\
  \caption{System model of a downlink underlay secure cognitive radio network}
  \vspace{-0.25cm}
\end{figure}
arbitrary positive semidefinite matrices, precoding designs can also be cast as GQMP problems.

In the sequel, we focus on two different transmission scenarios in the downlink underlay secure CR network: secure multicast scenario and secure broadcast scenario.
\subsection{Secure Multicast Scenario}
In the secure multicast scenario, the ST sends a common message to all SRs. Therefore, the channel input $\bs$ can be represented as
\begin{align}
\bs=\bP\bx
\end{align}
where $\bP\!\in\!\mathds{C}^{N_{\scriptscriptstyle T}\times N_{\scriptscriptstyle T}}$ is the precoding matrix at the ST, and $\bx\!\in\!\mathds{C}^{N_{\scriptscriptstyle T}\times 1}$ is the input data vector for SRs with zero-mean and covariance $E_{\bx}\big[\bx\bx^\mathrm{H}\big]\!=\!\bI$. We further assume that the $i$-th SR knows the instantaneous channel realization of $\bH_i$, the $j$-th ED knows the instantaneous channel realization of $\bG_{\!j}$, and the ST only has channel statistics of all nodes in the system, i.e., the transmit correlation matrices $\{\bm{\Theta}_{\!h_i},\bm{\Theta}_{\!g_j},\bm{\Theta}_{\!f_k}, \forall (i,j,k)\}$ as well as the distributions of $\tilde{\bH}_i$, $\tilde{\bG}_{\!j}$ and $\tilde{\bF}_{\!k}$. Then the average secrecy rate for the $i$-th SR can be expressed as
\begin{align}
R_i&=\min_{\substack{1\leq j\leq J}}\big[\mathcal{I}(\bx_i;\by_i|\bH_i)-\mathcal{I}(\bx_i;\bz_j|\bG_{\!j})\big]^{+}\label{M1}\\
&=\min_{\substack{1\leq j\leq J}}\big[E_{\scriptscriptstyle\bH_i}\mathcal{I}(\bx_i;\by_i|\bH_i\!=\!\mathbf{\bar{H}}_i)\nonumber\\
&\quad\quad\quad\quad\quad\quad -E_{\scriptscriptstyle\bG_{\!j}}\mathcal{I}(\bx_i;\bz_j|\bG_{\!j}\!=\!\mathbf{\bar{G}}_{\!j})\big]^{+}\label{M2}
\end{align}
where $\mathbf{\bar{H}}_i$ and $\mathbf{\bar{G}}_j$ are given channel realizations of $\bH_i$ and $\bG_{\!j}$, respectively. For notational simplicity, in the following, we omit the given channel realization condition in mutual information expressions. Based on \eqref{M2}, the following average secrecy rate for multicast scenario is achievable \cite{liang2009compound}
\begin{align}\label{OBJ11}
R_{\scriptscriptstyle \mathrm{MC}}(\bP)&=\min_{\substack{1\leq i\leq I\\1\leq j\leq J}}\big[E_{\scriptscriptstyle\bH_i}\mathcal{I}(\bx;\by_i)-E_{\scriptscriptstyle\bG_{j}}\mathcal{I}(\bx;\bz_j)\big]^{+}\nonumber\\
&=\Bigg[\min_{\substack{1\leq i\leq I\\1\leq j\leq J}}\big[E_{\scriptscriptstyle\bH_i}\mathcal{I}(\bx;\by_i)-E_{\scriptscriptstyle\bG_{j}}\mathcal{I}(\bx;\bz_j)\big]\Bigg]^{+}.
\end{align}

We maximize $R_{\scriptscriptstyle \mathrm{MC}}(\bP)$ under the power constraint at the ST and the interference threshold constraints at PRs. The average transmit power at the ST is constrained to $\gamma_0$:
\begin{align}\label{CTR1}
&E_{\bx}\trace\big(\bP\bx\bx^\mathrm{H}\bP^\mathrm{H}\big)=\trace(\bP^\mathrm{H}\bP)\leq\gamma_0
\end{align}
and the average interference power at the $k$-th PR is limited by $N_{\scriptscriptstyle T}\!\cdot\!\gamma_{k}$:
\begin{align}\label{CTR2}
E_{\bx,\scriptscriptstyle \bF_{\!k}} &\trace\big(\bF_{\!k}\bP\bx\bx^\mathrm{H}\bP^\mathrm{H}\bF_{\!k}^\mathrm{H}\big)\nonumber\\
&=E_{\scriptscriptstyle \tilde{\bF}_{\!k}}\trace\Big(\bP^\mathrm{H}(\bm{\Theta}^{\frac{1}{2}}_{\!f_k})^{\mathrm{H}}\tilde{\bF}_{\!k}^{\mathrm{H}}\tilde{\bF}_{\!k}\bm{\Theta}^{\frac{1}{2}}_{\!f_k}\bP\Big)\nonumber\\
&=N_{\scriptscriptstyle T}\!\cdot\!\trace\big(\bP^\mathrm{H}\bm{\Theta}_{\!f_k}\bP\big)\leq N_{\scriptscriptstyle T}\!\cdot\!\gamma_{k}, \; \forall k.
\end{align}
The second equality in \eqref{CTR2} holds because 1) the entries of $\tilde{\bF}_{\!k}$ are i.i.d. complex Gaussian variables with zero-mean and unit-variance; 2) $\tilde{\bF}_{\!k}$ and $\bx$ are independent. Since both \eqref{CTR1} and \eqref{CTR2} are quadratic matrix constraints, we can incorporate \eqref{CTR1} into \eqref{CTR2} by defining $\bm{\Theta}_{\!f_0}\!=\!\bI$. Then the set of all feasible precoding matrices is given by
\begin{align}
\mathcal{P}_{\scriptscriptstyle \mathrm{MC}}\!=\!\set[\Big]{\bP\given\trace\big(\bP^\mathrm{H}\bm{\Theta}_{\!f_k}\bP\big)\leq \gamma_k,\; k=0,1,...,K}
\end{align}
and the linear precoding problem under secure multicast scenario can be formulated as
\begin{equation}\label{OPT1}
\begin{aligned}
& \underset{\bP\in \mathcal{P}_{\scriptscriptstyle \mathrm{MC}}}{\mathrm{maximize}}
& & \min_{\substack{1\leq i\leq I\\1\leq j\leq J}}\big[E_{\scriptscriptstyle\bH_i}\mathcal{I}(\bx;\by_i)-E_{\scriptscriptstyle\bG_{j}}\mathcal{I}(\bx;\bz_j)\big].
\end{aligned}
\end{equation}
Here we remove the nonnegative operator $[\cdot]^+$ in $R_{\scriptscriptstyle \mathrm{MC}}(\bP)$ without loss of optimality because 1) the maximum average secrecy rate obtained by problem \eqref{OPT1} is nonnegative since we can always achieve a zero objective value with $\bP=\bm{0}$; 2) when the maximum average secrecy rate is positive, $[\cdot]^+$ has no effect on the optimal precoders.

\subsection{Secure Broadcast Scenario}
In the secure broadcast scenario, the ST sends a private message to each SR. Therefore, the channel input $\bs$ can be represented as
\begin{align}
\bs=\sum_{i=1}^{I}\bP_{\!i}\bx_i=\bP\bx
\end{align}
where $\bP_{\!i}\!\in\!\mathds{C}^{N_{\scriptscriptstyle T}\times N_{\scriptscriptstyle T}}$ is the precoding matrix for the $i$-th SR, $\bx_i\!\in\!\mathds{C}^{N_{\scriptscriptstyle T}\times 1}$ is the input data vector for the $i$-th SR with zero-mean and covariance $E_{\bx_i}\big[\bx_i\bx_i^\mathrm{H}\big]=\bI$, $\bP\!=\!\big[\bP_{\!1},\bP_{\!2},...,\bP_{\!\scriptscriptstyle I}\big]$ and $\bx=\big[\bx_1^\mathrm{H},\bx_2^\mathrm{H},...,\bx_{\scriptscriptstyle I}^\mathrm{H}\big]^\mathrm{H}$. We further assume that the $i$-th SR knows the instantaneous channel realization of $\bH_i$, the $j$-th ED knows the instantaneous channel realization of $\bG_{\!j}$, and the ST only has channel statistics of all nodes in the system. In addition, the $i$-th SR treats signals of other SRs as interference, and the $j$-th ED can at best decode the signal of the $i$-th SR while treating signals of other SRs as interference. Under these assumptions, the average secrecy rate for the $i$-th SR can be expressed as
\begin{align}\label{BroadRate}
R_i=\min_{1\leq j\leq J}\big[E_{\scriptscriptstyle\bH_i}\mathcal{I}(\bx_i;\by_{i})-E_{\scriptscriptstyle\bG_j}\mathcal{I}(\bx_i;\bz_j)\big]^{+}.
\end{align}
Based on \eqref{BroadRate}, the following average secrecy sum rate for broadcast scenario is achievable
\begin{align}\label{BroadRateC}
R_{\scriptscriptstyle \mathrm{BC}}(\bP)\!=\!\sum_{i=1}^{I}\min_{1\leq j\leq J}\big[E_{\scriptscriptstyle\bH_i}\mathcal{I}(\bx_i;\by_{i})\!-\!E_{\scriptscriptstyle\bG_j}\mathcal{I}(\bx_i;\bz_j)\big]^{+}.
\end{align}
The average secrecy sum rate maximization in secure broadcast scenario with power and interference threshold constraints can then be formulated as
\begin{equation}\label{OPT2}
\begin{aligned}
& \underset{\bP\in\mathcal{P}_{\scriptscriptstyle \mathrm{BC}}}{\mathrm{maximize}}\quad R_{\scriptscriptstyle \mathrm{BC}}(\bP)
\end{aligned}
\end{equation}
where the feasible set $\mathcal{P}_{\scriptscriptstyle \mathrm{BC}}$ can be obtained from \eqref{CTR1} and \eqref{CTR2} with $\bm{\Theta}_{\!f_0}\!=\!\bI$:
\begin{align}
\mathcal{P}_{\scriptscriptstyle \mathrm{BC}}\!=\!\set[\Big]{\bP\given\trace\big(\bP^\mathrm{H}\bm{\Theta}_{\!f_k}\bP\big)\leq \gamma_k,\; k=0,1,..., K}.
\end{align}

\section{Precoding Design By Generalized Quadratic Matrix Programming}
\subsection{Secure Multicast Scenario}
In this subsection, we solve problem \eqref{OPT1} under finite-alphabet inputs by GQMP. Instead of Gaussian inputs, we assume that the input data vector $\bx$ is uniformly distributed from $\mathcal{X}^{N_{\scriptscriptstyle T}}\!=\!\big\{\bx|x_k\!\in\!\mathcal{X},\forall k\big\}$, where $x_k$ is the $k$-th element of $\bx$, and $\mathcal{X}$ is a $Q$-ary discrete constellation set. Then the average constellation-constrained mutual information $E_{\scriptscriptstyle \bH_i}\mathcal{I}(\bx;\by_i)$ and $E_{\scriptscriptstyle \bG_j}\mathcal{I}(\bx;\bz_j)$ can be expressed as \cite{xiao2011globally}
\begin{align}
&E_{\scriptscriptstyle\bH_i}\mathcal{I}(\bx;\by_{i})\!=\!\log M
\!-\!\frac{1}{M}\!\!\sum_{m=1}^{M}\!E_{\scriptscriptstyle\bH_i,\bn_{h_i}}\!\bigg\{\!\log\!\sum_{n=1}^{M}\!e^{-h_{m,n}^i}\!\bigg\}\label{AMI1}\\
&E_{\scriptscriptstyle\bG_j}\mathcal{I}(\bx;\bz_{j})\!=\!\log M \!-\!\frac{1}{M}\!\!\sum_{m=1}^{M} \!E_{\scriptscriptstyle\bG_j,\bn_{g_j}}\!\bigg\{\!\log\!\sum_{n=1}^{M}\!e^{-g_{m,n}^j}\!\bigg\}\label{AMI2}
\end{align}
where $M$ is equal to $Q^{N_{\scriptscriptstyle T}}$; $h_{m,n}^i\!=\!\|\bH_i\bP(\bx_m\!-\!\bx_n)\!+\!\bn_{h_i}\|^2\!-\!\|\bn_{h_i}\|^2$ and $g_{m,n}^j\!=\!\|\bG_{\!j}\bP(\bx_m\!-\!\bx_n)+\bn_{g_j}\|^2-\|\bn_{g_j}\|^2$, with $\bx_m$ and $\bx_n$ representing input realizations from $\mathcal{X}^{N_{\scriptscriptstyle T}}$.

The average constellation-constrained mutual information is difficult to compute directly because both $E_{\scriptscriptstyle \bH_i}\mathcal{I}(\bx;\by_i)$ and $E_{\scriptscriptstyle \bG_j}\mathcal{I}(\bx;\bz_j)$ in \eqref{AMI1} and \eqref{AMI2} have no closed form expressions. Moreover, the gradients of $E_{\scriptscriptstyle \bH_i}\mathcal{I}(\bx;\by_i)$ and $E_{\scriptscriptstyle \bG_j}\mathcal{I}(\bx;\bz_j)$ also have no closed form expressions. Although we can use Monte Carlo method and numerical integral to estimate $E_{\scriptscriptstyle \bH_i}\mathcal{I}(\bx;\by_i)$ and $E_{\scriptscriptstyle \bG_j}\mathcal{I}(\bx;\bz_j)$ as well as their gradients, the computational complexity is prohibitively high especially when the dimensions of $\bH_i$ and $\bG_j$ are large.

This difficulty can be mitigated by introducing an accurate approximation of the average mutual information in doubly correlated fading channels \cite{zeng2012linear}. The average secrecy rate with finite-alphabet inputs can then be approximated as
\begin{align}\label{APP11}
\Bigg[\min_{\substack{1\leq i\leq I\\1\leq j\leq J}} &\big[g(\bP^\mathrm{H}\bm{\Theta}_{\!g_j}\bP;{N_{\!\scriptscriptstyle E}}) \!-\!g(\bP^\mathrm{H}\bm{\Theta}_{\!h_i}\bP;{N_{\!\scriptscriptstyle R}})\big]\Bigg]^+
\end{align}
where $g(\bW;N)$ with $N\geq 1$ is given below
\begin{align}\label{AMI}
g(\bW;N)=\frac{1}{M}\!\sum_{m=1}^{M}\log\sum_{n=1}^{M}\Big(1+\frac{1}{2}\be_{mn}^\mathrm{H}\bW\be_{mn}\Big)^{-N}.
\end{align}
The approximation in \eqref{APP11} is very accurate for arbitrary correlation matrices and precoders, and the computational complexity of \eqref{APP11} is several orders of magnitude lower than that of the original average secrecy rate \cite{zeng2012linear}.

Using \eqref{APP11} as an alterative to replace the average secrecy rate, problem \eqref{OPT1} can be approximated as
\begin{equation}\label{APP_OPT1}
\begin{aligned}
& \underset{\bP\in \mathcal{P}_{\scriptscriptstyle \mathrm{MC}}}{\mathrm{maximize}}\;\min_{\substack{1\leq i\leq I\\1\leq j\leq J}}g(\bP^\mathrm{H}\bm{\Theta}_{\!g_j}\!\bP;{N_{\!\scriptscriptstyle E}})\!-\!g(\bP^\mathrm{H}\bm{\Theta}_{\!h_i}\!\bP;{N_{\!\scriptscriptstyle R}}).
\end{aligned}
\end{equation}

The following proposition indicates that \eqref{APP_OPT1} is a GQMP problem with the min-rate utility.
\begin{proposition}
$g(\bW;N)$ is a convex and MNI function of $\bW$.
\end{proposition}
\begin{IEEEproof}
See Appendix B.
\end{IEEEproof}

Based on Proposition 4, $g(\bP^\mathrm{H}\bm{\Theta}_{\!g_j}\bP;{N_{\!\scriptscriptstyle E}})-g(\bP^\mathrm{H}\bm{\Theta}_{\!h_i}\bP;{N_{\!\scriptscriptstyle R}})$ is a generalized quadratic matrix function of $\bP$ for all $(i,j)$. Therefore, \eqref{APP_OPT1} is a special case of problem \eqref{GQMPE} and we can solve it efficiently by Algorithm 1.

\subsection{Secure Broadcast Scenario}
In this subsection, we solve problem \eqref{OPT2} under finite-alphabet inputs by GQMP. Instead of Gaussian inputs, we assume that the input data vector $\bx$ is uniformly distributed from $\mathcal{X}^{N_{\scriptscriptstyle T}I}\!=\!\big\{\bx|x_k\!\in\!\mathcal{X},\forall k\big\}$, where $x_k$ is the $k$-th element of $\bx$, and $\mathcal{X}$ is a $Q$-ary equiprobable discrete constellation set. Then the average constellation-constrained mutual information $E_{\scriptscriptstyle \bH_i}\mathcal{I}(\bx_i;\by_i)$ and $E_{\scriptscriptstyle \bG_j}\mathcal{I}(\bx_i;\bz_j)$ can be expressed as
\begin{align}
&E_{\scriptscriptstyle\bH_i}\mathcal{I}(\bx_i;\by_{i})=\frac{1}{M}\!\sum_{m=1}^{M} \!E_{\scriptscriptstyle\bH_i,\bn_{h_i}}\!\big\{\log(a_m^{ii})\big\}\label{EM1}\\
&E_{\scriptscriptstyle\bG_{\!j}}\mathcal{I}(\bx_i;\bz_{j})=\frac{1}{M}\!\sum_{m=1}^{M} \!E_{\scriptscriptstyle\bG_j,\bn_{g_j}}\!\big\{\log(b_m^{ij})\big\}\label{EM2}
\end{align}
with
\begin{align}
a_{m}^{ii}=\frac{\sum\limits_{n=1}^{M}e^{-\|\bH_i\bP\bI_i(\bx_m-\bx_n)+\bn_{h_i}\|^2+\|\bn_{h_i}\|^2}}{\sum\limits_{n=1}^{M}e^{-\|\bH_i\bP(\bx_m-\bx_n)+\bn_{h_i}\|^2+\|\bn_{h_i}\|^2}}\\
b_{m}^{ij}=\frac{\sum\limits_{n=1}^{M}e^{-\|\bG_j\bP\bI_i(\bx_m-\bx_n)+\bn_{g_j}\|^2+\|\bn_{g_j}\|^2}}{\sum\limits_{n=1}^{M}e^{-\|\bG_j\bP(\bx_m-\bx_n)+\bn_{g_j}\|^2+\|\bn_{g_j}\|^2}}
\end{align}
where $M$ is equal to $Q^{N_{\scriptscriptstyle T}I}$; $\bI_i$ is a block diagonal matrix formed by replacing the $i$-th $N_{\scriptscriptstyle T}\times N_{\scriptscriptstyle T}$ block diagonal entry of the $N_{\scriptscriptstyle T}I\times N_{\scriptscriptstyle T}I$ identity matrix $\bI$ with $\mathbf{0}$; $\bx_m$ and $\bx_n$ are input realizations from $\mathcal{X}^{N_{\scriptscriptstyle T}I}$.

Similarly, by adopting the accurate approximation in \cite{zeng2012linear}, the average secrecy sum rate with finite-alphabet inputs can be approximated respectively as
\begin{align}\label{app_broadrate}
\sum_{i=1}^{I}\min_{1\leq j\leq J} \big[\mathcal{I}_{\scriptscriptstyle A}(\bx_i;\by_i)- \mathcal{I}_{\scriptscriptstyle A}(\bx_i;\bz_j)\big]^{+}
\end{align}
where $\mathcal{I}_{\scriptscriptstyle A}(\bx_i;\by_i)$ and $\mathcal{I}_{\scriptscriptstyle A}(\bx_i;\bz_j)$ are accurate approximations of $E_{\scriptscriptstyle\bH_i}\mathcal{I}(\bx_i;\by_{i})$ and $E_{\scriptscriptstyle\bG_j}\mathcal{I}(\bx_i;\bz_{j})$ respectively:
\begin{align}
&\mathcal{I}_{\scriptscriptstyle A}(\bx_i;\by_i)=g_i(\bP^\mathrm{H}\bm{\Theta}_{\!h_i}\bP;N_{\!\scriptscriptstyle R})-g(\bP^\mathrm{H}\bm{\Theta}_{\!h_i}\bP;N_{\!\scriptscriptstyle R})\\
&\mathcal{I}_{\scriptscriptstyle A}(\bx_i;\bz_j)=g_i(\bP^\mathrm{H}\bm{\Theta}_{\!g_j}\bP;N_{\!\scriptscriptstyle E})- g(\bP^\mathrm{H}\bm{\Theta}_{\!g_j}\bP;N_{\!\scriptscriptstyle E})
\end{align}
with
\begin{align}
&g_i(\bW;N)\!=\!\frac{1}{M}\!\sum_{m=1}^{M}\log\sum_{n=1}^{M}\Big(1\!+\!\frac{1}{2}\be_{mn}^\mathrm{H}\bI_i^\mathrm{H}\bW\bI_i\be_{mn}\Big)^{\!-N}\\
&g(\bW;N)\!=\!\frac{1}{M}\!\sum_{m=1}^{M}\log\sum_{n=1}^{M}\Big(1\!+\!\frac{1}{2}\be_{mn}^\mathrm{H}\bW\be_{mn}\Big)^{\!-N}.
\end{align}
Using \eqref{app_broadrate} as an alternative, problem \eqref{OPT2} can be approximated as
\begin{equation}\label{APD_BC}
\begin{aligned}
& \underset{\bP\in \mathcal{P}_{\scriptscriptstyle \mathrm{BC}}}{\mathrm{maximize}}
& & \sum_{i=1}^{I}\min_{1\leq j\leq J} \big[\mathcal{I}_{\scriptscriptstyle A}(\bx_i;\by_i)- \mathcal{I}_{\scriptscriptstyle A}(\bx_i;\bz_j)\big]^{+}
\end{aligned}
\end{equation}

According to Proposition 4, $g_i(\bW;N)$ and $g(\bW;N)$ are convex and MNI functions of $\bW$. Therefore, problem \eqref{APD_BC} is a special case of problem \eqref{GGQMP}, and we can solve it efficiently by Algorithm 2.

\subsection{Discussions}
In this subsection, we show that precoding problems under the doubly correlated model \eqref{CMOD} are GQMP problems. For convenience, we restrict our attention to the secure multicast scenario, and it is straightforward to extend our results to the secure broadcast scenario.

Our first step is to replace the average secrecy rate $R_{\scriptscriptstyle\mathrm{MC}}(\bP)$ by the following accurate approximation:
\begin{align}\label{AASS}
\min_{\substack{1\leq i\leq I\\1\leq j\leq J}} &\Big[\bar{g}\big(\bP^\mathrm{H}\bm{\Theta}_{\!g_j}\bP;\bm{\lambda}(\bm{\Phi}_{\!g_j})\big) \!-\!\bar{g}\big(\bP^\mathrm{H}\bm{\Theta}_{\!h_i}\bP;\bm{\lambda}(\bm{\Phi}_{\!h_i})\big)\Big]^+
\end{align}
where $\bm{\lambda}(\cdot)$ represents the eigenvalues of a positive semidefinite matrix; $\bar{g}(\bW;\br)$ is given below \cite{zeng2012linear}
\begin{align}
\bar{g}(\bW;\br)\!=\!\frac{1}{M}\!\sum_{m=1}^{M} \log\sum_{n=1}^{M}\prod_q\Big(1+\frac{r_q}{2}\be_{mn}^\mathrm{H}\bW\be_{mn}\Big)^{-1}
\end{align}
with $r_q\!\geq\! 0$ being the $q$-th element of $\br$. When the receive correlation matrix is an identity matrix $\bI$, $\bm{\lambda}(\bI)$ is the vector with all entries one, and $\bar{g}(\bW;\bm{\lambda}(\bI))$ is reduced to $g(\bW;N)$ in \eqref{AMI}. Although $\bar{g}(\bW;\br)$ looks more complicated than $g(\bW;N)$, the convexity and MNI property of $\bar{g}(\bW;\br)$ can be proved in the same manner as in Proposition 4. This leads to the following proposition.
\begin{proposition}
$\bar{g}(\bW;\br)$ is a convex and MNI function of $\bW$.
\end{proposition}
\begin{IEEEproof}
The proof is omitted for brevity.
\end{IEEEproof}

The second step is to determine the interference threshold constraints at PRs. When $\bm{\Phi}_{\!f_k}\neq \bI$, the average interference power at the $k$-th PR is given by
\begin{align}
E_{\bx,\scriptscriptstyle \bF_{\!k}} &\trace\big(\bF_{\!k}\bP\bx\bx^\mathrm{H}\bP^\mathrm{H}\bF_{\!k}^\mathrm{H}\big)\nonumber\\
&\!=\!E_{\scriptscriptstyle \tilde{\bF}_{\!k}}\trace\Big(\bP^\mathrm{H}(\bm{\Theta}^{\frac{1}{2}}_{\!f_k})^{\mathrm{H}}\tilde{\bF}_{\!k}^{\mathrm{H}}\bm{\Phi}_{\!f_k}\tilde{\bF}_{\!k}\bm{\Theta}^{\frac{1}{2}}_{\!f_k}\bP\Big)\nonumber\\
&\!=\!\trace(\bm{\Phi}_{\!f_k})\!\cdot\!\trace\big(\bP^\mathrm{H}\bm{\Theta}_{\!f_k}\bP\big).
\end{align}
\begin{figure}[ht]
  \begin{center}
  \includegraphics[scale=0.55]{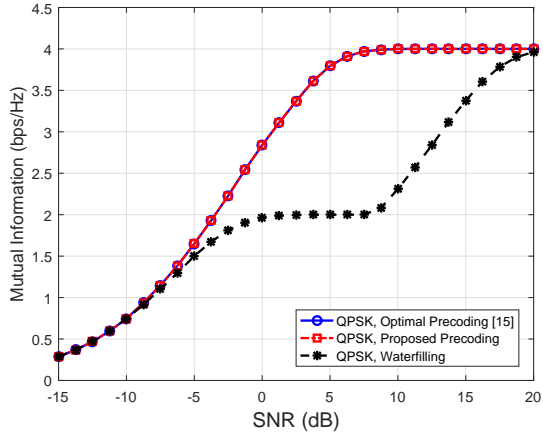}\\
  \caption{Mutual information as a function of the SNR.}
  \end{center}
  \vspace{-0.5cm}
\end{figure}
Since $\trace(\bm{\Phi}_{\!f_k})>0$ and $\bm{\Theta}_{\!f_k}\succeq \bm{0}$, the average interference power is a convex function of $\bP$.

According to Proposition 5, the approximated average secrecy rate is a generalized quadratic matrix function with the min-rate utility. In addition, the feasible set with power and interference threshold constraints is convex. Therefore, the precoding problem under secure multicast scenario is a GQMP problem, and it can be solved efficiently by Algorithm 1.

\section{Numerical Results}
In this section, we provide numerical examples to show that the GQMP algorithm for precoding design is numerically robust and computationally effective under various situations. For illustration purpose, we adopt the exponential correlation model
\begin{align} \label{CorrM}
\big[\bR(\rho)\big]_{ij}=\rho^{|i-j|}, \quad \rho\in[0,1)
\end{align}
where $\rho$ is the correlation coefficient between different antennas.

\subsection{Example 1: Point-to-point MIMO Gaussian Channel}
In this example, we compare our proposed precoding with the globally optimal precoding algorithm \cite{xiao2011globally} in a point-to-point MIMO setting. Specifically, we solve problem \eqref{PPPA} with finite-alphabet inputs by the GQMP and the globally optimal precoding algorithms. The channel matrix $\bH$ is given by
\begin{align}
\bH\!=\!
\begin{bmatrix}
    2       & 1 \\
    1       & 1
\end{bmatrix}
\end{align}
which was also used in \cite{xiao2011globally}. The input signal is drawn from QPSK constellation. Since we assume unit noise power, the signal-to-noise ratio (SNR) is defined as $\mathrm{SNR}=\gamma$, where $\gamma$ is the total transmit power at the transmitter.

Fig. 2 depicts comparison results with the globally optimal precoding algorithm \cite{xiao2011globally} and the waterfilling algorithm. As shown in Fig. 2, our proposed precoding can achieve the performance of the globally optimal precoding algorithm in whole SNR regimes. In addition, the waterfilling algorithm, which is optimal for Gaussian inputs, is quite suboptimal for practical MIMO systems under finite-alphabet inputs, especially in the medium and high SNR regimes.

\subsection{Example 2: MIMO Gaussian Wiretap Channel}
In this example, we compare our proposed precoding with the gradient descent algorithm \cite{wu2012linearw} in a MIMO wiretap channel. Specifically, we solve problem \eqref{WC} with finite-alphabet inputs by the GQMP and the gradient descent algorithms. The channel matrices at the receiver and the eavesdropper are given respectively as
\begin{align}
&\bH_\mathrm{r}\!=\!
\begin{bmatrix}
    0.0991-0.8676i & 1.0814 + 1.1281i
\end{bmatrix},\\
&\bH_\mathrm{e}\!=\!
\begin{bmatrix}
0.3880 + 1.2024i& -0.9825 + 0.5914i\\
0.4709 - 0.3073i& 0.6815 - 0.2125i
\end{bmatrix}
\end{align}
which was also used in \cite{wu2012linearw}. The input signal is drawn from QPSK constellation, and the SNR is defined as $\mathrm{SNR}=\frac{\gamma}{\sigma^2}$, where $\sigma^2$ is the noise power at the receiver and the eavesdropper. In addition, the initial point for both algorithms is set as
\begin{align}
\bP_{\!0}\!=\!
\begin{bmatrix}
0.0312 - 0.1762i   &0.1719 + 0.7560i\\
   0.9126 + 0.5724i & -0.1064 - 0.0097i
\end{bmatrix}.
\end{align}

\begin{figure}[ht]
  \begin{center}
  \includegraphics[scale=0.55]{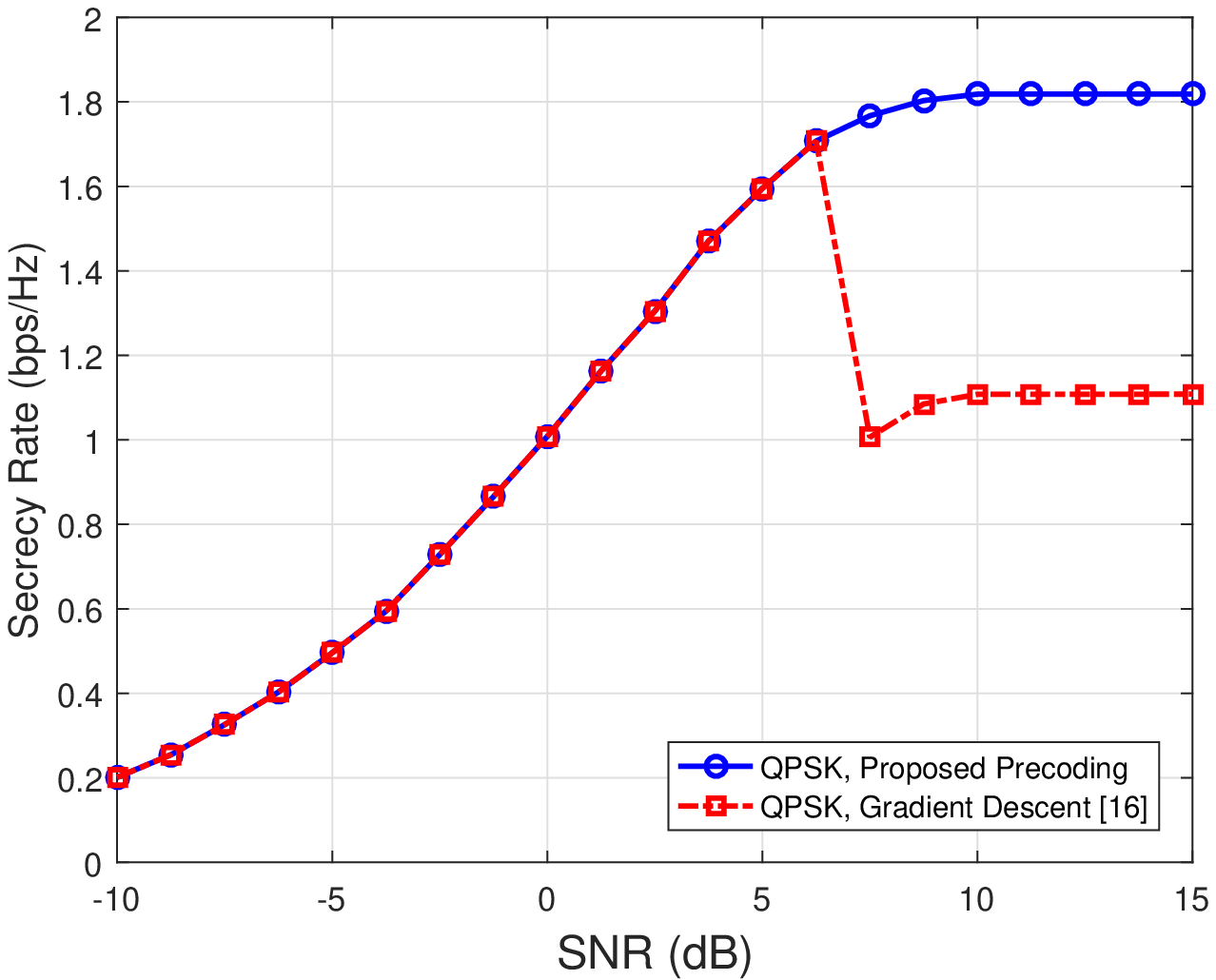}\\
  \caption{Comparison between the proposed precoding and the gradient descent algorithms with finite-alphabet inputs.}
  \end{center}
  \vspace{-0.9cm}
\end{figure}
Fig. 3 compares the secrecy rate performance of our proposed precoding with the gradient descent algorithm \cite{wu2012linearw}. We observe that our algorithm outperforms the gradient descent algorithm in high SNR regimes because the gradient descent algorithm is susceptible to initial points.

To further show the influence of initial points for the proposed precoding and the gradient descent algorithms, we run these two algorithms 2000 times with Gaussian distributed initial points (applying a normalization to satisfy the power constraint).

\begin{table}[!ht]
\centering
\begin{tabular}{c c c}
Secrecy Rate& Proposed Precoding & Gradient Descent \\
\hline\hline
1.7471 bps/Hz & 1831/2000 & 932/2000\\
1.0063 bps/Hz & 169/2000 & 1068/2000\\
\hline
\end{tabular}
\vspace{0.1cm}
\caption{Distribution of secrecy rate for various initial points.}
\vspace{-0.4cm}
\end{table}

Table 1 shows the distribution of secrecy rate for the proposed precoding and the gradient descent algorithm with $\mathrm{SNR}\!=\!7.5\mathrm{dB}$. The result demonstrates the superiority of the proposed algorithm. Our algorithm has over $90\%$ probability to converge to a better solution while the gradient descent algorithm only has $46.6\%$ probability to achieve this goal.

\subsection{Example 3: Secure Multicast Scenario with One ED}
In this example, we consider a fading secure CR network with one ST, two SRs, one ED and one PR. Each node in the system has two antennas. The transmit correlation matrices are given by
\begin{align}
&\bm{\Theta}_{\!h_1}=\bR(0.95), \bm{\Theta}_{\!h_2}=\bR(0.85),\nonumber\\
&\bm{\Theta}_{\!g_1}=\bR(\rho), \bm{\Theta}_{\!f_1}=\bR(0.50).
\end{align}
The SNR is defined as $\mathrm{SNR}=\gamma_0$, where $\gamma_0$ is the total transmit power at the ST. The normalized interference threshold at the PR is 10 dB less than the total transmit power, i.e., $\gamma_1=0.1\gamma_0$.

\begin{figure}[ht]
  \begin{center}
  \includegraphics[scale=0.55]{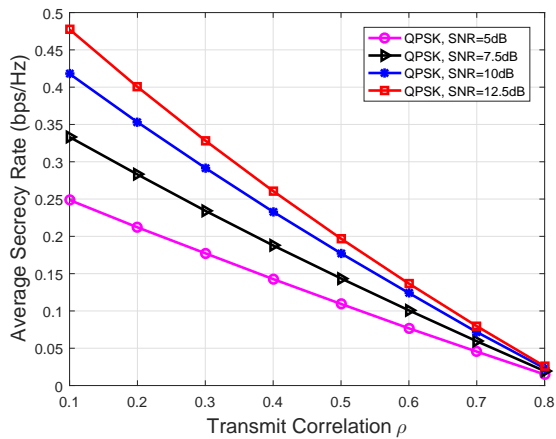}\\
  \caption{Average secrecy rate as a function of the ED's transmit correlation $\rho$.}
  \end{center}
  \vspace{-0.5cm}
\end{figure}

Fig. 4 investigates the maximum average secrecy rate as a function of the ED's transmit correlation $\rho$ with $\mathrm{SNR}\!=\!5 \mathrm{dB}$, $7.5 \mathrm{dB}$, $10 \mathrm{dB}$ and $12.5 \mathrm{dB}$. The input signal is drawn from QPSK constellation. As we can see, the average secrecy rate is monotonically decreasing with respect to $\rho$. This interesting phenomenon occurs because the impact of transmit correlation depends on the channel knowledge \cite{jorswieck2004optimal}. Reference \cite{jorswieck2004optimal} considered a single-user channel and showed that if the transmitter only knows statistical CSI, the average capacity is Schur-convex with respect to the channel correlation, i.e., the more correlated the transmit antennas are, the more capacity can be achieved. Therefore, when $\rho$ increases, we need more redundancy rate to confuse the ED and the average secrecy rate decreases. In the extreme case that $\rho=0.85$, the average secrecy rate is always zero because $\bm{\Theta}_{\!h_2}=\bm{\Theta}_{\!g_1}$.

In Fig. 5, we investigate the average secrecy rate as a function of the SNR under BPSK, QPSK, 8PSK and 16QAM modulations. The ED's transmit correlation is set as $\rho=0.2$. Fig. 5 shows that precoding design by the proposed GQMP algorithm is very effective because it can achieve robust performances for a large-SNR range with various modulations. These results also indicate that we should adaptively determine the modulation based on the SNR. If the system works in the low SNR regime, we should use low-order modulations to reduce the complexity for precoder design. If the system works in the high SNR regime, we can use high-order modulations to achieve a better performance.

\subsection{Example 4: Secure Multicast Scenario with Multiple EDs}
In this example, we consider a fading secure CR network with one ST, two SRs, two EDs and two PRs. Each node in the system has two antennas. The transmit correlation matrices are given by
\begin{align}
&\bm{\Theta}_{\!h_1}=\bR(0.9), \bm{\Theta}_{\!h_2}=\bR(0.95), \bm{\Theta}_{\!g_1}=\bR(0.3)\nonumber \\
&\bm{\Theta}_{\!g_2}=\bR(0.4), \bm{\Theta}_{\!f_1}=\bR(0.5), \bm{\Theta}_{\!f_2}=\bR(0.7).
\end{align}
The normalized interference thresholds at PRs are 20dB less than the total transmit power, i.e., $\gamma_1=\gamma_2=0.01\gamma_0$, and the SNR is defined as $\mathrm{SNR}=\gamma_0$.

Fig. 6 depicts the comparison result with the Gaussian precoding design under QPSK and 16QAM modulations. The Gaussian precoding design employs Gaussian inputs to solve problem \eqref{OPT1},
\begin{figure}[ht]
  \begin{center}
  \includegraphics[scale=0.55]{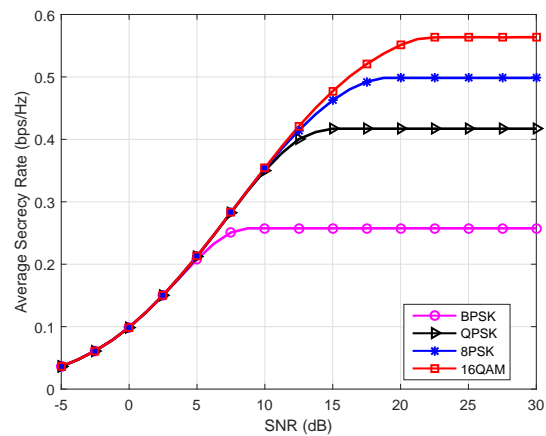}\\
  \caption{Average secrecy rate as a function of the SNR.}
  \end{center}
  \vspace{-0.5cm}
\end{figure}
where $E_{\scriptscriptstyle\bH_i}\mathcal{I}(\bx;\by_i)$ and $E_{\scriptscriptstyle\bG_j}\mathcal{I}(\bx;\bz_j)$ with Gaussian inputs can be expressed as
\begin{align}
&E_{\scriptscriptstyle\bH_i}\mathcal{I}(\bx;\by_i)=E_{\scriptscriptstyle\bH_i}\Big[\log\det\big(\bI+\bH_i\bP\bP^\mathrm{H}\bH_i^\mathrm{H}\big)\Big]\\
&E_{\scriptscriptstyle\bG_j}\mathcal{I}(\bx;\bz_j)=E_{\scriptscriptstyle\bG_{j}}\Big[\log\det\big(\bI+\bG_{\!j}\bP\bP^\mathrm{H}\bG_{\!j}^\mathrm{H}\big)\Big].
\end{align}
Let $\bQ=\bP\bP^\mathrm{H}$, then the Gaussian precoding problem is given by
\begin{equation}\label{GP}
\begin{aligned}
& \underset{\bQ\succeq \bm{0}}{\mathrm{maximize}}
& & \min_{\substack{1\leq i\leq I\\1\leq j\leq J}}\!\big[E_{\scriptscriptstyle\bH_i}\mathcal{I}(\bx;\by_i)\!-\!E_{\scriptscriptstyle\bG_j}\mathcal{I}(\bx;\bz_j)\big]\\
& \mathrm{subject \;to}
& & \quad\trace(\bm{\Theta}_{\!f_k}\!\bQ)\!\leq\! \gamma_k, \; k=0,1,...,K
\end{aligned}
\end{equation}

Since problem \eqref{GP} is a DC problem, we can obtain a locally optimal solution $\bQ_{\mathrm{opt}}$ by the convex-concave procedure \cite{yuille2003concave}. After that, we evaluate the finite-alphabet based average secrecy rate under $\bP=\bQ_{\mathrm{opt}}^{\frac{1}{2}}$.

Based on the results in Fig. 6, we have the following remarks:

1) In the low SNR regime, finite-alphabet precoding and Gaussian precoding have the same performance. According to \cite{perez2010mimo}, the low-SNR expansion of mutual information is irrelevant to the input distribution, thus the optimal precoders designed under Gaussian inputs are also optimal for finite-alphabet inputs.

2) In the high SNR regime, the performance of Gaussian precoding design degrades severely with the increasing SNR. The reason is that both $E_{\scriptscriptstyle \bH_i}\mathcal{I}(\bx;\by_i)$ and $E_{\scriptscriptstyle \bG_j}\mathcal{I}(\bx;\bz_j)$ in \eqref{AMI1}, \eqref{AMI2} saturate at $\log M$ in the high SNR regime. Therefore, if we do not carefully control the transmit power $\trace(\bP^\mathrm{H}\bP)$ in the high SNR, the average secrecy rate under finite-alphabet inputs will be $\log M-\log M=0$. Since the objective function of \eqref{GP} does not have this saturation property, the covariance matrix $\bQ_{\mathrm{opt}}$ designed by problem \eqref{GP} uses too much transmit power. Therefore, the corresponding average secrecy rate with finite-alphabet inputs degrades severely in the high SNR regime.

\subsection{Example 5: Secure Broadcast Scenario with Multiple EDs}
In this example, we consider a fading secure CR network with one ST, two SRs, two EDs and two PRs. Each node
\begin{figure}[ht]
  \begin{center}
  \includegraphics[scale=0.55]{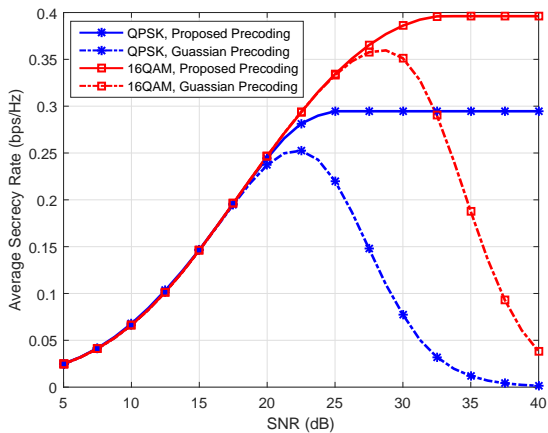}\\
  \caption{Proposed precoding versus Gaussian precoding design under different modulations, for secure multicast scenarios.}
  \end{center}
  \vspace{-0.5cm}
\end{figure}
in the system has two antennas. The transmit correlation matrices are given by
\begin{align}
&\bm{\Theta}_{\!h_1}=\bR(0.9), \bm{\Theta}_{\!h_2}=\bR(0.8), \bm{\Theta}_{\!g_1}=\bR(0.45),\nonumber\\
&\bm{\Theta}_{\!g_2}=\bR(0.55), \bm{\Theta}_{\!f_1}=\bR(0.2), \bm{\Theta}_{\!f_2}=\bR(0.6).
\end{align}
The normalized interference thresholds at PRs are 10dB less than the total transmit power, i.e., $\gamma_1=\gamma_2=0.1\gamma_0$, and the SNR is defined as $\mathrm{SNR}=\gamma_0$.

Fig. 7 depicts the performance comparison with the Gaussian precoding design under BPSK and QPSK modulations. The Gaussian precoding solves problem \eqref{OPT2} under Gaussian inputs by GQMP, and then evaluates the finite-alphabet based average secrecy sum rate with the corresponding suboptimal precoders. Results in Fig. 6 indicate that our proposed finite-alphabet precoding offers much higher secrecy sum rate than the Gaussian precoding in the medium and high SNR regimes. This is because the precoders $[\bP_{\! 1}, \bP_{\! 2}]$ designed by Gaussian precoding have the following form:
\begin{align}\label{GPC}
[\bP_{\! 1}, \bP_{\! 2}]=[\bP_{\! 1}^{\mathrm{opt}}, \mathbf{0}].
\end{align}
Equation \eqref{GPC} implies that the Gaussian precoding design allocates all the power to the first SR, and the precoder for the second SR is $\mathbf{0}$. In contrast, our proposed finite-alphabet precoding allocates power to both SRs, and the precoded signal for the second SR acts as the jamming signal to further confuse EDs. Therefore, our proposed precoding significantly outperforms the Gaussian precoding design.

\section{Conclusion}
In this paper, we have proposed the generalized quadratic matrix programming, which is a significant generalization of quadratic matrix programming. GQMP captures the inherent structure of input-output mutual information with arbitrary input distributions, thus it unifies the design of linear precoding under various MIMO Gaussian channels. By exploiting the features of generalized quadratic matrix functions, we have designed a low complexity algorithm that converges to the KKT point of any GQMP problem. Next, we have applied GQMP to design linear precoders in downlink
\begin{figure}[ht]
  \begin{center}
  \includegraphics[scale=0.55]{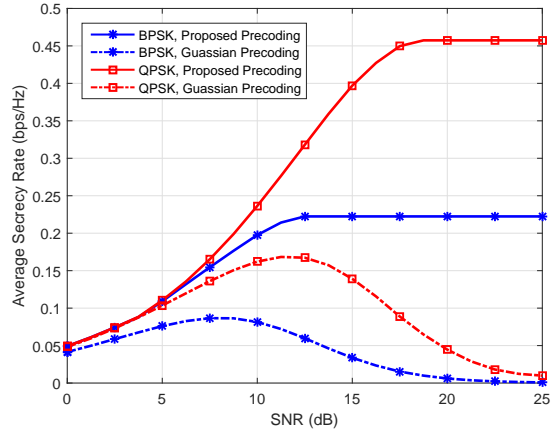}\\
  \caption{Proposed precoding versus Gaussian precoding design under different modulations, for secure broadcast scenarios.}
  \end{center}
  \vspace{-0.5cm}
\end{figure}
underlay secure CR networks. The considered linear precoding target finite-alphabet inputs directly and exploit statistical CSI of fading channels. We have demonstrated that the precoding problems under both secure multicast and secure broadcast scenarios are GQMP problems, thus we can solve them efficiently by our proposed GQMP algorithm. Numerical results have shown that the proposed algorithm is both robust and effective.

\section*{Appendix A\\ Proofs of Theorems 1 and 2}
\begin{IEEEproof}[Proof of Theorem 1]
Since $g(\bW)$ is a differentiable convex function, it can be lower bounded by its first-order Taylor approximation at any point $\bW_{\!0}$ \cite[ch. 3.1.3]{boyd2004convex}:
\begin{align}\label{T11}
g(\bW)\geq g(\bW_{\!0})+\trace\big[(\bW-\bW_{\!0})^\mathrm{H}\nabla_{\!\scriptscriptstyle\bW}g(\bW_{\!0})\big].
\end{align}
By plugging $\bW\!=\!\bX^\mathrm{H}\bA\bX$, $\bW_{\!0}\!=\!\bX_0^\mathrm{H}\bA\bX_0$ and $\bG\!=\!\nabla_{\!\scriptscriptstyle\bW}g(\bW_{\!0})$ into \eqref{T11}, we have
\begin{align}\label{T12}
g(\bX&^\mathrm{H}\bA\bX)\nonumber\\
&\geq \trace\big(\bX^\mathrm{H}\bA\bX\bG\big)\!+\!g(\bX_0^\mathrm{H}\bA\bX_0)\!-\!\trace\big(\bX_0^\mathrm{H}\bA\bX_0\bG\big).
\end{align}
Since $\trace(\bY^\mathrm{H}\bV\bY\bZ)=\mathrm{vec}(\bY)^\mathrm{H}\cdot\big(\bZ^{\mathrm{T}} \otimes \bV\big)\cdot \mathrm{vec}(\bY)$ \cite{horn1994matrix}, we can express $\trace\big(\bX^\mathrm{H}\bA\bX\bG\big)$ alternatively as
\begin{align}
&\trace\big(\bX^\mathrm{H}\bA\bX\bG\big)=\trace\big(\bX^\mathrm{H}\bA^{\!\scriptscriptstyle(+)}\bX\bG\big)+\trace\big(\bX^\mathrm{H}\bA^{\!\scriptscriptstyle(-)}\bX\bG\big) \label{T13}\\
&=\mathrm{vec}(\bX)^\mathrm{H}\cdot\Big[\big(\bG^{\mathrm{T}} \otimes \bA^{\!\scriptscriptstyle(+)}\big)\!+\!\big(\bG^{\mathrm{T}} \otimes \bA^{\!\scriptscriptstyle(-)}\big)\Big]\cdot \mathrm{vec}(\bX).
\end{align}

If $g(\bW)$ is MND, its complex gradient $\bG\succeq \bm{0}$ for all $\bX_0$\cite[ch. 3.6.1]{boyd2004convex}. Therefore, $\bG^{\mathrm{T}} \otimes \bA^{\!\scriptscriptstyle(+)}\succeq \bm{0}$ and $\bG^{\mathrm{T}} \otimes \bA^{\!\scriptscriptstyle(-)}\preceq\!\bm{0}$ \cite{horn1994matrix}, which imply the convexity of $\trace\big(\bX^\mathrm{H}\bA^{\!\scriptscriptstyle(+)}\bX\bG\big)$ and the concavity of $\trace\big(\bX^\mathrm{H}\bA^{\!\scriptscriptstyle(-)}\bX\bG\big)$ respectively.
The convex function $\trace\big(\bX^\mathrm{H}\bA^{\!\scriptscriptstyle(+)}\bX\bG\big)$ can be further lower bounded by its first-order Taylor approximation at $\bX_0$:
\begin{align}
&\trace\big(\bX^\mathrm{H}\bA^{\!\scriptscriptstyle(+)}\bX\bG\big)\nonumber\\
&\quad\geq \trace\big(\bX^\mathrm{H}\bA^{\!\scriptscriptstyle(+)}\bX_0+\bX_0^\mathrm{H}\bA^{\!\scriptscriptstyle(+)}\bX
-\bX_0^\mathrm{H}\bA^{\!\scriptscriptstyle(+)}\bX_0\bG\big). \label{T14}
\end{align}
Combining \eqref{T12}, \eqref{T13} and \eqref{T14}, we obtain a concave lower bound when $g(\bW)$ is MND.

If $g(\bW)$ is MNI, its complex gradient $\bG\preceq \bm{0}$ for all $\bX_0$ \cite[ch. 3.6.1]{boyd2004convex}. Therefore, $\bG^{\mathrm{T}} \otimes \bA^{\!\scriptscriptstyle(+)}\preceq \bm{0}$ and $\bG^{\mathrm{T}} \otimes \bA^{\!\scriptscriptstyle(-)}\succeq\!\bm{0}$ \cite{horn1994matrix}, which imply the concavity of $\trace\big(\bX^\mathrm{H}\bA^{\!\scriptscriptstyle(+)}\bX\bG\big)$ and the convexity of $\trace\big(\bX^\mathrm{H}\bA^{\!\scriptscriptstyle(-)}\bX\bG\big)$ respectively. The convex function $\trace\big(\bX^\mathrm{H}\bA^{\!\scriptscriptstyle(-)}\bX\bG\big)$ can be further lower bounded by its first-order Taylor approximation at $\bX_0$:
\begin{align}
&\trace\big(\bX^\mathrm{H}\bA^{\!\scriptscriptstyle(-)}\bX\bG\big)\nonumber\\
&\quad\geq \trace\big(\bX^\mathrm{H}\bA^{\!\scriptscriptstyle(-)}\bX_0+\bX_0^\mathrm{H}\bA^{\!\scriptscriptstyle(-)}\bX
-\bX_0^\mathrm{H}\bA^{\!\scriptscriptstyle(-)}\bX_0\bG\big). \label{T15}
\end{align}
Combining \eqref{T12}, \eqref{T13} and \eqref{T15}, we obtain a concave lower bound when $g(\bW)$ is MNI.
\end{IEEEproof}

\begin{IEEEproof}[Proof of Theorem 2]
Since $(\bX-\bX_0)^\mathrm{H}\bA^{\!\scriptscriptstyle(-)}(\bX-\bX_0)\preceq \bm{0}$ and $(\bX-\bX_0)^\mathrm{H}\bA^{\!\scriptscriptstyle(+)}(\bX-\bX_0)\succeq \bm{0}$, the following inequalities hold
\begin{align}
&\bX^\mathrm{H}\bA\bX \preceq \bU_1(\bX)\\
&\bX^\mathrm{H}\bA\bX \succeq \bU_2(\bX)
\end{align}
with
\begin{align*}
&\bU_1(\bX)\!=\!\bX^\mathrm{H}\bA^{\!\scriptscriptstyle(+)}\bX\!+\!\bX^\mathrm{H}\bA^{\!\scriptscriptstyle(-)}\bX_0\!+\!\bX_0^\mathrm{H}\bA^{\!\scriptscriptstyle(-)}\bX\!-\!\bX_0^\mathrm{H}\bA^{\!\scriptscriptstyle(-)}\bX_0\\ &\bU_2(\bX)\!=\!\bX^\mathrm{H}\bA^{\!\scriptscriptstyle(-)}\bX\!+\!\bX^\mathrm{H}\bA^{\!\scriptscriptstyle(+)}\bX_0\!+\!\bX_0^\mathrm{H}\bA^{\!\scriptscriptstyle(+)}\bX\!-\!\bX_0^\mathrm{H}\bA^{\!\scriptscriptstyle(+)}\bX_0.
\end{align*}
Here $\bU_1(\bX)$ is a convex function in the sense that for all $\bX_1$, $\bX_2$ and $\theta$ with $0\leq\theta\leq1$, we have
\begin{align}
\bU_1(\theta\bX_1\!+\!(1\!-\!\theta)\bX_2)\preceq \theta \bU_1(\bX_1)+(1\!-\!\theta) \bU_1(\bX_2). \label{T21}
\end{align}
Based on the definition in \eqref{T21}, $-\bU_2(\bX)$ is convex, thus $\bU_2(\bX)$ is a concave function.

If $g(\bW)$ is MND, $g\big(\bU_1(\bX)\big)$ serves as an upper bound of $g(\bX^\mathrm{H}\bA\bX)$. Furthermore, for all $\bX_1$, $\bX_2$ and $\theta$ with $0\leq\theta\leq1$, we have
\begin{align}
g\big(\bU_1&(\theta\bX_1+(1-\theta)\bX_2)\big)\nonumber\\
&\leq g\big(\theta \bU_1(\bX_1)+(1-\theta) \bU_1(\bX_2)\big) \label{T22}\\
&\leq \theta g\big(\bU_1(\bX_1)\big)+(1-\theta) g\big(\bU_1(\bX_2)\big)\label{T23}
\end{align}
where \eqref{T22} and \eqref{T23} hold due to the MND condition and the convexity of $g(\bW)$, respectively. Therefore, $g\big(\bU_1(\bX)\big)$ serves as a convex upper bound of $g(\bX^\mathrm{H}\bA\bX)$ when $g(\bW)$ is MND.

If $g(\bW)$ is MNI, $g\big(\bU_2(\bX)\big)$ serves as an upper bound of $g(\bX^\mathrm{H}\bA\bX)$. Furthermore, for all $\bX_1$, $\bX_2$ and $\theta$ with $0\leq\theta\leq1$, we have
\begin{align}
g\big(\bU_2&(\theta\bX_1+(1-\theta)\bX_2)\big)\nonumber\\
&\leq g\big(\theta \bU_2(\bX_1)+(1-\theta) \bU_2(\bX_2)\big) \label{T24}\\
&\leq \theta g\big(\bU_2(\bX_1)\big)+(1-\theta) g\big(\bU_2(\bX_2)\big)\label{T25}
\end{align}
where \eqref{T24} and \eqref{T25} hold due to the MNI condition and the convexity of $g(\bW)$, respectively. Therefore, $g\big(\bU_2(\bX)\big)$ serves as a convex upper bound of $g(\bX^\mathrm{H}\bA\bX)$ when $g(\bW)$ is MNI.
\end{IEEEproof}

\section*{Appendix B\\ Proofs of Propositions 1--4}
\begin{IEEEproof}[Proof of Proposition 1]
First, by introducing an auxiliary variable $t$, problem \eqref{GQMP} is equivalent to
\begin{equation}\label{RS}
\begin{aligned}
& \underset{\bX\in \mathcal{X},t}{\mathrm{maximize}}
& & t\\
& \mathrm{subject \;to}
& & f_0(\bX)-t\geq 0\\
&&& f_j(\bX)\geq 0, \; j=1,2,...,J.
\end{aligned}
\end{equation}
Then we can directly apply the inner approximation method in \cite{marks1978technical} to problem \eqref{RS}. In each iteration, we keep the objective function $t$ unchanged and approximate all constraints at $\bX_0$:
\begin{equation}\label{RS1}
\begin{aligned}
& \underset{\bX\in \mathcal{X},t}{\mathrm{maximize}}
& & t\\
& \mathrm{subject \;to}
& & \bar{f}_0(\bX;\bX_0)-t\geq 0\\
&&& \bar{f}_j(\bX;\bX_0)\geq 0, \; j=1,2,...,J
\end{aligned}
\end{equation}
where $\bar{f}_j(\bX;\bX_0)$ are the corresponding concave lower bound of $f_j(\bX)$. Note that problem \eqref{RS1} is equivalent to problem \eqref{GQMPLB}, which is the subproblem in each iteration of Algorithm 1. Therefore, Algorithm 1 and the proposed algorithm in \cite{marks1978technical} are equivalent. Based on Theorem 1 in \cite{marks1978technical}, Algorithm 1 converges to the KKT point of problem \eqref{GQMP}. This completes the proof.
\end{IEEEproof}

\begin{IEEEproof}[Proof of Proposition 2]
Recall that problem \eqref{GGQMP} is
\begin{equation}\label{R_1231}
\begin{aligned}
R=& \underset{\bX\in \mathcal{X}}{\mathrm{maximize}}
& & \sum_{i=1}^{I}\big[F_i(\bX)\big]^{+}
\end{aligned}
\end{equation}
where $F_i(\bX)$ is given by
\begin{align}
F_i(\bX)=\min_{1\leq j\leq J} f_{ij}(\bX).
\end{align}
Define the set of nonnegative $F_i(\bX)$ at the optimal solution of problem \eqref{R_1231} as $\mathcal{S}=\big\{i|F_i(\bX_{\mathrm{opt}})\geq 0, i=1,2,...,I\big\}$, where $\bX_{\mathrm{opt}}$ denotes the optimal solution of \eqref{R_1231}. Then problem \eqref{R_1231} is equivalent to
\begin{equation}\label{R_12311}
\begin{aligned}
& \underset{\bX\in \mathcal{X}}{\mathrm{maximize}}
& & \sum_{i\in \mathcal{S}} F_i(\bX).
\end{aligned}
\end{equation}
Note that $\mathcal{S}$ can be any non-empty subset of $\{1,2,...,I\}$, i.e., $\mathcal{S}\!\in\! \{\mathcal{S}_m,m\!=\!1,2,...,2^I\!-1\}$, where $\mathcal{S}_m$ is the $m$-th non-empty subset of $\{1,2,...,I\}$. Therefore, we need to solve problem \eqref{R_12311} $2^I-1$ times with different $\mathcal{S}_m$ to obtain $R$. This completes the proof.
\end{IEEEproof}

\begin{IEEEproof}[Proof of Proposition 3]
We first show that $\{F(\bX_{n}^{\mathrm{opt}})\}$ is monotonically nondecreasing by the following inequalities:
\begin{align}
F(\bX_{n+1}^{\mathrm{opt}})\geq\bar{F}(\bX_{n+1}^{\mathrm{opt}};\bX_{n}^{\mathrm{opt}})\geq\bar{F}(\bX_{n}^{\mathrm{opt}};\bX_{n}^{\mathrm{opt}})=F(\bX_{n}^{\mathrm{opt}})
\end{align}
where the first inequality holds because $\bar{F}(\cdot;\bX_{n}^{\mathrm{opt}})$ serves as the lower bound of $F(\cdot)$; the second inequality holds because $\bX_{n+1}^{\mathrm{opt}}$ is the optimal solution of $\underset{\bX\in \mathcal{X}}{\mathrm{maximize}} \; \bar{F}(\bX;\bX_{n}^{\mathrm{opt}})$. Furthermore, since the feasible set $\mathcal{X}$ is compact and $F(\cdot)$ is continuous, the sequence $\{F(\bX_{n}^{\mathrm{opt}})\}$ is bounded above. Therefore, the convergence is guaranteed since every bounded monotone sequence has a limit. This completes the proof.
\end{IEEEproof}

\begin{IEEEproof}[Proof of Proposition 4]
The convexity can be shown by reformulating $g(\bW;N)$ as
\begin{align}
\frac{1}{M}\sum_{m=1}^{M} \log\sum_{n=1}^{M}\exp\Big[\!-\!N\!\cdot\!\ln\big(1+\frac{1}{2}\be_{mn}^\mathrm{H}\bW\be_{mn}\big)\Big].
\end{align}
Since $-\ln\big(1+0.5\cdot\be_{mn}^\mathrm{H}\bW\be_{mn}\big)$ is a convex function of $\bW$ for any vector $\be_{mn}$, and $\log\sum_{i}\exp(f_i)$ is convex whenever $f_i$ is convex for all $i$, $g(\bW;N)$ is a convex function.

The MNI property can be shown by computing the complex gradient of $g(\bW;N)$
\begin{align}
\nabla_{\!\scriptscriptstyle \bW}g(\bW,N)=-\sum_{m,n}\alpha_{mn}\be_{mn}\be_{mn}^\mathrm{H}
\end{align}
where
\begin{align}
\alpha_{mn}=\frac{N}{2M}\cdot \frac{\big(1\!+\!0.5\cdot\be_{mn}^\mathrm{H}\!\bW\be_{mn}\big)^{-\!N\!-\!1}}{\sum_{n=1}^{M}\big(1\!+\!0.5\cdot\be_{mn}^\mathrm{H}\!\bW\be_{mn}\big)^{-\!N}}>0.
\end{align}
Since $\nabla_{\!\scriptscriptstyle \bW}g(\bW,N)\preceq \bm{0}$, $g(\bW,N)$ is a MNI function \cite[ch. 3.6.1]{boyd2004convex}. This completes the proof.
\end{IEEEproof}
\maketitle
\bibliographystyle{IEEEtran}
\bibliography{IEEEabrv,reference}

\end{document}